\documentclass{article}
\usepackage{graphicx}
\usepackage{mathptmx}
\usepackage{amsmath}
\usepackage[all]{xy}
\usepackage{amssymb}
\usepackage{url}

\newtheorem{thm}{Theorem}
\newtheorem{cor}[thm]{Corollary}
\newtheorem{lem}[thm]{Lemma}

\newtheorem{dfn}[thm]{Definition}

\newtheorem{rem}[thm]{Remark}

\newcommand{\qed}{\hfill \hbox{\rule[-2pt]{3pt}{6pt}} \par}

\newcommand{\E}{\mathcal{E}}

\newcommand{\hil}{\mathcal{H}}
\newcommand{\B}{\mathcal{B}}
\newcommand{\complex}{\mathbb{C}}
\newcommand{\trace}[1][]{{\rm Tr}[{#1}]}
\newcommand{\state}{\mathcal{S}}

\newcommand{\proj}{\Pi}

\newcommand{\iden}{\mathbb{I}}

\newcommand{\id}{{\rm id}}

\newcommand{\hilb}[1][]{\mathcal{H}_{#1}}

\newcommand{\ketbra}[1][]{| #1 \rangle \langle #1 |}
\newcommand{\ket}[1][]{| #1 \rangle}

\newcommand{\topos}{\mathcal{T}}
\newcommand{\sets}{\mathbf{Sets}}
\newcommand{\ccstar}{\mathbf{cCstar}}

\newcommand{\R}{\mathbb{R}}

\newcommand{\tr}{{\rm Tr}}
\newcommand{\mH}{H}

\newcommand{\mW}{\mathcal{W}}
\newcommand{\Hi}{\mathcal{H}_1 \otimes \cdots \otimes \mathcal{H}_n}
\newcommand{\gA}[1][]{\underline{\Sigma}_{A_{#1}}}

\newcommand{\gAi}{\underline{\Sigma}_{A_1} \times \cdots \times \underline{\Sigma}_{A_n}}

\newcommand{\uR}{\underline{\mathbb{R}}}
\newcommand{\uA}[1][]{\underline{A_{#1}}}

\newcommand{\uAi}{\underline{A_1} \otimes \cdots \otimes \underline{A_n}}
\newcommand{\cAi}{{\com}(A_1) \times \cdots \times {\com}(A_n)}

\newcommand{\com}{{\mathcal{C}}}
\newcommand{\topAi}{[ \prod_{i=1}^n {\com}(A_i),\, \mathbf{Set}]}

\newcommand{\loc}{\mathbf{Loc}}

\newcommand{\category}{\mathbf{C}}
\newcommand{\T}{\mathbf{T}}
\newcommand{\dist}{\mathcal{D}}

\newcommand{\kl}{{\mathcal{K}\ell}}
\newcommand{\func}{T}
\newcommand{\monad}{(\func, \mu, \eta)}

\newcommand{\UFF}{{\rm UFF}^1({H}_1, \ldots, {H}_n)}
\newcommand{\integ}{\mathcal{I}}
\newcommand{\val}{\mathcal{V}}
\newcommand{\str}{\mathrm{st}}
\newcommand{\cst}{\mathrm{cst}}
\newcommand{\dst}{\mathrm{dst}}
\newcommand{\ext}{\mathrm{ext}}

\newcommand{\Hom}{{\rm Hom}}

\newcommand{\cont}{{\rm Cont}}
\newcommand{\fub}{i}

\begin{document}

\title{Composite systems and state transformations in topos quantum theory}

\author{{Jisho Miyazaki} \\[2mm]
	\small Department of Physics, Graduate School of Science, University of Tokyo, Japan}

\date{\today}
\maketitle

\begin{abstract}
Topos quantum theory provides representations of quantum states as direct generalizations of the probability distribution, namely probability valuation.
In this article, we consider extensions of a known bijective correspondence between quantum states and probability valuations to composite systems and to state transformations.
We show that multipartite probability valuations on composite systems have a bijective correspondence to positive over pure tensor states, according to a candidate definition of the composite systems in topos quantum theory.
Among the multipartite probability valuations, a special attention is placed to Markov chains which are defined by generalizing classical Markov chains from probability theory.
We find an incompatibility between the multipartite probability valuations and a monogamy property of quantum states, which trivializes the Markov chains to product probability valuations.
Several observations on the transformations of probability valuations are deduced from the results on multipartite probability valuations, through duality relations between multipartite states and state transformations.
\end{abstract}

\section{Introduction}
Topos quantum theory (see e.g.~\cite{DoringIsham2011,Wolters2013} for reviews) provides a representation of quantum states as probability weight functions that generalize classical probability distributions, much like quasi-probability distributions \cite{Lee1995:quasiprobinfinite,FerrieEmerson2009:framedspace} and Gleason's measures \cite{Gleason1975}.
In topos quantum theory, the space of random variables and probability distributions are generalized to \emph{locales} and \emph{valuations}, respectively.
Major branches of topos quantum theories usually start from finding appropriate topos for a given Hilbert space or C*-algebra representing the quantum system of interest, and then internal locales are constructed so that there exists a bijective correspondence between quantum states on the original system and valuations on the locale \cite{DoringIsham2008I,HeunenLandsmanSpitters2009}.

As represented by the negative values of Wigner functions, any probability weight function expressing a quantum state differs from a mere probability distribution.
In topos quantum theory, the difference between quantum states and probability distributions appears especially in the locales.
The internal locales for representing quantum states do not usually have points \cite{IshamButterfield1998,HeunenLandsmanSpitters2009}, while the spaces of random variables always do (see, e.g.~\cite{Johnstone1986:stone} for the definition of points on locales, and \cite{Vickers2007:localetoposasspaces} for their roles in logic).
Points of the locale, if exist, reveal non-contextual value assignments to physical observables, which is shown impossible for quantum theory by Kochen-Specker theorem \cite{KochenSpecker1975}.
Topos quantum theories connect probability distributions and quantum states seamlessly, by employing pointless locales to avoid contradiction to the Kochen-Specker theorem.

Although the bijective correspondence is shown to exist between valuations on suitable locales and quantum states, we still have only a limited understanding on the properties of valuations inherited from probability distributions and quantum states.
In particular, valuations have to abandon one of the properties from quantum states and probability distributions, when these two conflict to each other.
In this case valuations have to abandon the property originated from probability distributions to be a suitable representation of quantum states, much like they drop points of the locale to exhibit the contextuality.
In this article, we place a special attention on the behaviors of valuations on composite systems and transformations on valuations as presented in Table\,\ref{tab:TQT and probability theory}, and investigate similarities and differences between their counterparts in the usual quantum theory.
\begin{table}
	\caption{Analogous concepts in classical probability theory, topos quantum theory, and quantum theory.	TPCP stands for ``trace preserving and completely positive''.}
	\label{tab:TQT and probability theory}
	\begin{tabular}{llll}
		\hline\noalign{\smallskip}
		& probability theory & topos quantum theory & quantum theory \\
		\noalign{\smallskip}\hline\noalign{\smallskip}
		system & random variable (set) & locale & Hilbert spaces \\
		states & probability distribution & valuation & density operators \\
		composites & product & Sec.\,\ref{chap:composite topos} & tensor product \\
		transformations & transition matrix & Sec.\,\ref{chap:Markov} & TPCP maps \\
		\noalign{\smallskip}\hline
	\end{tabular}
\end{table}

Composite systems in topos quantum theory remain so unexplored that currently one cannot find a canonical definition of composite systems, unlike the tensor products of Hilbert spaces for quantum theory and products of random variables for probability theory.
There are analyses on independence conditions of local systems \cite{Nuiten2011,WoltersHalvorson2013} in one of the major branches of topos quantum theory called the Bohrification approach \cite{HeunenLandsmanSpitters2009} (also called ``Nijmegen approach''), and one can find a candidate of composite systems in Ref.\,\cite{WoltersHalvorson2013} although their motivation is different from defining composite systems.
Since we do not have the canonical definition, we employ the candidate definition of the composite systems, and proceed the analysis further to the valuations therein.
In particular, we investigate how the bijective correspondence between quantum states and valuations, studied so far only on uncomposable single systems, generalizes to the composite systems.
This step would show us whether the multipartite properties of quantum states such as entanglement and non-locality can be inherited to valuations, and may provide insights on how to improve the definition of composite systems, if necessary.

Transformations on valuations in topos quantum theory remain also unexplored, but this time, a result from constructive analysis suggests a canonical definition of the transformations.
Recall that transition matrices are Kleilsi morphisms of the distribution monad (see e.g.\,\cite{Jacobs2011:distributionmonad}).
Kleilsi morphisms of monads are used to represent more general probabilistic transformations \cite{Giry1982,FurberJacobs2015,Westerbaan2017}.
In \cite{Vickers2011:monad}, Vickers worked on valuations, partly motivated from topos quantum theory, and had shown that valuations also constitutes of a commutative monad.
The canonical transformations on valuations would be given by the Kleisli morphisms of the valuation monad, which generalize transition matrices.
We apply Vickers' results to the Bohrification approach, and investigate how the bijective correspondence between valuations and quantum states generalizes to their respective transformations (see the last line of Table\,\ref{tab:TQT and probability theory}).

Since the Kleisli morphisms between valuations are difficult to handle compared to the valuations themselves, we instead consider a generalization of classical Markov chains to the Bohrification approach. 
Markov chains in classical probability theory are a particular kind of joint probability distributions on composite systems \cite{CoverThomas2006}, and have already been generalized to quantum theory in several ways \cite{Rivas2014:markov,BuscemiDatta2016,BaeChuscinski2016,Petz1986,Petz1988,Accardi1982:quantumstochastic,Hayden2004:ssa}.
These Markov chains share a property that long Markov chains are constructed from short Markov chains by extending the latter with certain transformations such as transition matrices and completely positive maps.
In any cases, Markov chains reflect properties of state transformations from which they are constructed.

We start from the analysis on the composite systems, and show a bijective correspondence between the set of valuations on the composite systems and the set of positive over pure tensor states \cite{Barnum2010:popt} under a restriction on the dimension of the Hilbert space of interest.
Positive over pure tensor states are equivalent to usual quantum states on single systems, but include operators that differs from the quantum states on composite systems.
Therefore we recover the original bijective correspondence shown in Ref.\,\cite{HeunenLandsmanSpitters2009} for single systems, but this correspondence is not generalized straightforwardly to the composite systems.
The gap between valuations and quantum states on composite systems has an implication on the definition of complete positivity of the Kleisli morphisms of the valuation monad, through a duality relation between bipartite operators and state transformations.

We then generalize Markov chains to the Bohrification approach.
The defined Markov chains are certain valuations on composite locales constructed from consecutive actions of Kleisli morphisms of the valuation monad, and thus reflect properties of these transformations.
We show several properties shared by Markov chains of classical probability theory and of the Bohrification approach.
We find, however, an incompatibility between these shared properties and the monogamy of quantum states that trivializes Markov chains in the Bohrification approach to product states.
Several observations on the Kelisli morphisms are deduced from this result.  

This article is organized as follows.
We give a short introduction of topos quantum theory with emphasis on its relation to classical probability theory in Sec.\,\ref{sec:intrototopos}.
In Sec.\,\ref{chap:composite topos} we define composite systems in topos quantum theory as a certain product of locales representing marginal systems, and show a bijective correspondence between the valuations on the composite systems and positive over pure tensor states.
In Sec.\,\ref{sec:markov chains for monads}, we leave topos for a moment and formulate Markov chains for general commutative monads on cartesian categories.
Based on the results obtained in Secs.\,\ref{chap:composite topos} and \ref{sec:markov chains for monads}, we define Markov chains in topos quantum theory, analyze their properties, and make several observations on the Kleisli morphisms of the valuation monad in Sec.\,\ref{chap:Markov}.
Finally, we lay our conclusion in Sec.\,\ref{chap:conclusion}.

\section{Introduction to topos quantum theory}\label{sec:intrototopos}
There are several branches in topos quantum theory \cite{DoringIsham2008I,HeunenLandsmanSpitters2009,Wolters2013}.	A basic idea common to all of them is to find quantum theory as a direct generalization of classical probability theory by using topos.	From a topos theory point of view, classical probability theory is a topos quantum theory formalized in the particluar topos $\sets$.

We briefly recall classical probability theory before moving on to topos.	A random variable is represented by a finite set, say, $X$.	A probability distribution $p$ on $X$ is a function $p:X \to [0,1]$ such that $\sum_{x \in X} p(x) = 1$.	Equivalently, a probability distribution is an assignment of probability weights on measurable subsets of $X$.	If the set of subsets of the finite set $X$ is denoted by ${\rm Sub}(X)$, the assignment is a function $p:{\rm Sub}(X) \to [0,1]$ satisfying conditions
\begin{eqnarray}
&& p(X)=1,~p(\emptyset)=0, \label{eq:measure1}\\
&& p(V) \leq p(W) \text{ if } V \subset W \text{ in } {\rm Sub}(X),\label{eq:measure3}\\
&& p(V) + p(W) = p(V \cap W) + p(V \cup W),\label{eq:measure4}
\end{eqnarray}
where $W$ and $V$ are subsets of $X$.\footnote{If we consider probability distributions as probability measures on discrete sets, more natural choice of conditions would be Eq.\,(\ref{eq:measure1}) and the $\sigma$-additivity.	The $\sigma$-additivity is equivalent to conditions (\ref{eq:measure3}) and (\ref{eq:measure4}).} 

Topos quantum theory uses probability \emph{valuations} to represent quantum states (we shall omit adjective ``probability'' since all valuations appearing in this thesis is normalized).	Valuations are similar to probability measures, but differs especially in that they are defined on \emph{locales}.	A locale $X$ is a certain lattice whose elements are called ``opens''. If a locale further has a certain structures of opens called ``points,'' it is called a topological space.	The set of subset of $X$, $\mathrm{Sub}(X)$ forms a topological space whose points corresponds to the elements of $X$.	Locales in general do not necessarily have points.	A valuation on a locale is an assignment of probability weights on the opens of the locale such that the conditions analogous to Eqs.\,(\ref{eq:measure1}), (\ref{eq:measure3}) and (\ref{eq:measure4}) are satisfied (see Sec.\,\ref{ssec:Gelfand and Riesz} for details).

Topos quantum theory differs from classical probability theory not only in the use of locales and valuations but also in the category on which they are defined.	For defining concepts and proving theorems in classical probability theory, we usually use set-theoretic logic.	Toposes are categories with associated languages which enables interpretation of formal logical expressions inside the toposes (see, e.g.~\cite{MacLaneMoerdijk1992} for details).	From topos theoretic point of view, set-theoretic logic is the language associated to topos $\sets$.	We shall call mathematical concepts to be \emph{internal} to a topos, if they are interpreted by the language of the topos.	

Major branches of topos quantum theories \cite{DoringIsham2008I,HeunenLandsmanSpitters2009} start from finding proper toposes for a given Hilbert space $\hil$ (or more generally, a non-commutative C*-algebra).	 Then internal locales are constructed so that valuations on them have one-to-one correspondence to quantum states on $\hil$.	We obtain the representation of quantum states by valuations in this way.

Among several branches of topos quantum theory, we focus on one of the mainstreams called ``Bohrification approach'' \cite{HeunenLandsmanSpitters2009} (also called ``Nijmegen approach'').	The Bohrification approach makes full use of the languages associated to topos.	Although the axiom of choice and the law of excluded middle are no longer valid in a general topos, any mathematical theorems proven \emph{constructively} are interpreted to be valid in any topos.	Constructively proven theorems contribute to showing the existence of the set of valuations corresponding to quantum states.	Since proposers of the other mainstream ``Imperial approach'' \cite{DoringIsham2008I} do not make use of the internal language of toposes, constructively proven theorems are not necessarily interpretable by their approach.	Our result presented in this article are not applicable for the Imperial approach since we use several known theorems from constructive mathematics.	We give detailed introductions to these constructively proven theorems in Sec.\,\ref{ssec:Gelfand and Riesz}, and to the Bohrification approach in Sec.\,\ref{ssec:Bohrification}.

\subsection{Constructive Gelfand duality and Riesz theorem}\label{ssec:Gelfand and Riesz}
In this subsection, we introduce two known theorems from constructive mathematics which play significant roles in Sec.\,\ref{chap:composite topos}.	The Gelfand duality is a celebrated result in theory of algebras.	Classically it gives a duality between commutative C*-algebras and compact Hausdorff spaces both internal to $\sets$, and constructive proofs extended the duality between general commutative C*-algebras and compact, completely regular locales.	The Riesz theorem (also called Riesz-Markov theorem) expresses another duality relation between integrals over commutative C*-algebras and valuations on locales.	

The constructive Gelfand duality is represented by a contravariant equivalence between categories of commutative C*-algebras $\mathbf{cCstar}$ and compact, completely regular locales $\mathbf{KCRegLoc}$.	The proof of Gelfand duality to be constructive implies that this contravariant equivalence holds in any toposes.	More precisely, we have following theorem:
\begin{thm}\label{thm:Gelfand duality}
	Let $\topos$ be any topos and $\mathbf{cCstar}$ and $\mathbf{KCRegLoc}$ be categories of commutative C*-algebras and compact, completely regular locales defined internally to $\topos$.	There is a pair of contravariant functors $\Sigma:\mathbf{cCstar} \to \mathbf{KCRegLoc}$ and $\cont (-,\complex):\mathbf{KCRegLoc} \to \mathbf{cCstar}$ such that $\Sigma \circ \cont (-,\complex)$ and $\cont (-,\complex) \circ \Sigma$ are the identity functors on $\mathbf{KCRegLoc}$ and $\mathbf{cCstar}$, respectively.
\end{thm}
\noindent This theorem is implied by the results presented in Ref.\,\cite{BanaschewskiMulvey2006}, and summarized in this form in Ref.\,\cite{HeunenLandsmanSpitters2009}.	In the following argument we do not specify the topos in question.

The compact completely regular locale $\Sigma_A := \Sigma(A)$ for an internal commutative C$^*$-algebra $A$ is called the \emph{Gelfand spectrum} for $A$.	Although there is a method to construct $\Sigma_A$ from $A$, we can refer to the valuations on $\Sigma_A$ without specifying the precise form of $\Sigma_A$.	This is possible by combining the Gelfand duality and the Riesz theorem presenting a duality between following valuations and integrals.

A \emph{valuation} on locale $X$ is a morphism $v:\mathcal{O}(X) \to [0,1]$ such that
\begin{eqnarray}
&& v (\bot) = 0,\\
&& v(\top) = 1,\\
&& v(U) + v(V) = v(U \vee V) + v(U \wedge V), \\
&& v(U) \leq v(V) \text{  if  } U \leq V \text{  in  } X
\end{eqnarray}
for any opens $U,~V$ of $X$, where $\bot$ and $\top$ represents the bottom and top elements of $X$, respectively\footnote{Precisely speaking, the morphism $v$ is required to satisfy the following continuity: $\sup v(V_i) = v(\sup V_i)$ for directed family $\{ V_i \}$.	A family of opens $\{ V_i \}$ is called directed if for any pair $V,W \in \{ V_i \}$, there exists $X \in \{ V_i \}$ such that $V \leq X$ and $W \leq V$.}.	These conditions must be interpreted by internal languages of toposes.

An \emph{integral} over a self-adjoint part $C_{sa}$ of a C$^*$-algebra $C$ (we call it an integral over $C$ for short) is formally defined as a morphism $I:C_{sa} \rightarrow \R$ such that
\begin{eqnarray}
\nonumber I(\iden) = 1 && \hspace{1cm} \textrm{(normalization)},\\
\nonumber I(a + b) = I(a) + I(b) && \hspace{1cm} \textrm{(linearlity)},\\
\label{eq:positivity of integral} I(a) \geq 0 \text{ if } a \geq 0&& \hspace{1cm} \textrm{(positivity)}
\end{eqnarray}
hold for all $a,b \in C_{sa}$ and $\iden$, the identity in $C_{sa}$	\cite{CoquandSpitters2009:integralvaluation} \footnote{Most generally integrals are defined on f-algebras \cite{CoquandSpitters2009:integralvaluation}.	Here we only consider those f-algebras equivalent to self-adjoint parts of C$^*$-algebras.}.	The conditions presented above should be interpreted by internal languages of toposes.

The constructive Riesz theorem states that there is a bijective correspondence between integrals over commutative C$^*$-algebra $A$ and valuations on the Gelfand spectrum $\Sigma_A$.	Given an integral $I$, we can construct an associated valuation $\mu_I$, and vice versa.

The constructive Riesz theorem states the equivalence not only of the integrals and valuations themselves, but also of the corresponding locales.	Let $1$ be the terminal object of internal locales.	For each internal commutative C$^*$-algebra $A$, there is an associated locale $\integ A$ such that the set $\Hom_\loc (1,\integ A)$ has a bijective correspondence to integrals over $A$ ($\Hom$ refers to the set of morphisms). For each internal locale $X$, there is an associated locale $\val X$ such that the set $\Hom_\loc (1,\val X)$ has a bijective correspondence to valuations on $X$.	The constructive Riesz theorem presented in Ref.\,\cite{CoquandSpitters2009:integralvaluation} states the existence of the isomorphism
\begin{eqnarray}
\integ A \cong \val \Sigma_A,
\label{eq:constructive Riesz theorem}
\end{eqnarray}
in the category of locales, if commutative C$^*$-algebra $A$ and locale $\Sigma_A$ are Gelfand dual.	This implies an equivalence between $\Hom_\loc (1,\integ A)$ (integrals over $A$) and $\Hom_\loc (1,\val \Sigma_A)$ (valuations on $\Sigma_A$).	We do not write down the precise definitions of these locales here because it requires familiarity on logics (for interested readers, we recommend Ref.\,\cite{Vickers2007:localetoposasspaces}), and because we only use several of their known properties which could be stated without their definitions.

Now the Gelfand duality and the Riezs theorem can be schematically summarized by the following diagram:
\begin{equation}\label{eq:constructive Gelfand and Riesz}
	\begin{aligned}\xymatrix{A \text{ in } \mathbf{cCstar} \ar@<0.5ex>[d]^{\Sigma} \ar@{.>}[r]_{\integ} & \integ A \text{ in } \loc \ar@{=}[d]^\sim & \Hom_\loc (1,\integ A) \cong \text{Integrals over }A \ar@{=}[d]^\sim \\
		\Sigma_A \text{ in } \mathbf{KCRegLoc} \ar@<0.5ex>[u]^{\cont (-,\complex)} \ar@{.>}[r]_{\val} & \val \Sigma_A \text{ in } \loc. & \Hom_\loc (1,\val \Sigma_A) \cong \text{Valuations on }\Sigma_A
	}\end{aligned}
\end{equation}
The isomorphism (\ref{eq:constructive Riesz theorem}) provides a way to analyze states in two ways.	In this section, we show the one-to-one correspondence between positive over pure tensor states and integrals over algebras on composite systems.	The isomorphism (\ref{eq:constructive Riesz theorem}) then implies the one-to-one correspondence between positive over pure tensor states and valuations.

\subsection{Bohrification approach}\label{ssec:Bohrification}
In this subsection, we give a more detailed introduction to the Bohrification approach first presented in Ref.\,\cite{HeunenLandsmanSpitters2009}.	There are several branches inside the Bohrification approach \cite{Raynaund2014,Henry2015:geometricbohr}.	Among these sub-branches, we use the original approach for two reasons.	First, the correspondence between quantum states and integrals or valuations has been most explicitly investigated in the original approach.	Second, the internal language has a relatively simple form in the covariant functor topos $[P, \sets]$ over poset $P$ used in the original approach.	For these two reasons, the original approach is suitable for starting the investigation of the composite systems.
Since our investigation is restricted to finite dimensional quantum systems, \cite{Caspers2009:nlevel} also provides a sufficient background.

Let $A$ be a non-cmmutative algebra $\B(\hil)$ of bounded operators of a finite dimensional quantum system $\hil$.
Let $\com(A)$ be a poset of unital \emph{commutative} subalgebras of $A$, ordered by the subalgebra inclusion, i.e.~$C \leq D$ if and only if $C \subset D$.	Each commutative C*-subalgebra $C$ of a non-commutative C*-algebra $A$, namely, each element of poset $\com(A)$ is refered to as a ``context''.

The covariant approach uses the covariant functor topos $[\com(A),\sets]$ for describing the quantum system $A$.	The objects of $[\com(A),\sets]$ are functors from poset $\com(A)$ to $\sets$, and morphisms are natural transformations between these functors.

The {\it Bohrification} of algebra $A$ is the unital commutative C$^*-$algebra object $\uA$ internal to $[\com(A),\sets]$ defined by
\begin{eqnarray*}
\uA (C) = C,
\end{eqnarray*}
for each unital comutative C$^*-$algebra $C \in \com(A)$.	A unital commutative C$^*-$algebra object in topos $\topos$ is the object which satisfy axioms of unital commutative C$^*-$algebras interpreted by the internal language of topos $\topos$ (see e.g.~\cite{MacLaneMoerdijk1992} for details).

It is shown in Ref.\,\cite{HeunenLandsmanSpitters2009} that there is a bijective correspondence between the set of integrals over $\uA$ internal to $[\com(A),\sets]$, and the set of quasi-states on $A$, if the dimension of $A$ is greater than $2$.	If $A = \B(\hil)$ for a finite dimensional Hilbert space $\hil$ ($\dim \hil \geq 3$), quasi-states on $A$ are equivalent to density operators\footnote{A quasi-state on $A$ is a map $\rho: A \rightarrow \complex$ that is positive and linear on all commutative subalgebras and satisfies $\rho (a+ib) = \rho(a) + i\rho(b)$ for all self-adjoint $a, b \in A $ (possibly non-commuting). If $A$ is a von Neumann algebra and does not contain a type-II von Neumann factor, quasi-states are just usual quantum states.	In particular, $A$ does not contain type-II von Neumann factor if $A=\B(\hil)$ for certain finite dimensional Hilbert space $\hil$.}.	If the conditions for integrals are interpreted in $[\com(A),\sets]$, an integral over $\uA$ is a natural transformation $I:\uA_{sa} \to \uR$, where $\uR$ is the real number object in $[\com(A),\sets]$, such that each component $I_C:C_{sa} \to \R$ ($C \in \com(A)$) is an integral in $\sets$ \cite{HeunenLandsmanSpitters2009}.	Given a density operator $\rho$, we can define the corresponding integral $I_\rho$ by $I_{\rho C} (a):=\trace[\rho a]$ ($\forall C \in \com(A)$, $\forall a \in C_{sa}$).	Conversely, given an integral $I:\uA_{sa} \to \uR$, we can define the corresponding density matrix $\rho_I$ by $\trace[\rho a] := I_C (a)$ ($\forall a \in A_{sa}$), where $C \in \com(A)$ is any context such that $a \in C_{sa}$.	Once we have the equivalence between density operators and integrals, the equivalence between density operators and valuations immediately follows from the isomorphism (\ref{eq:constructive Riesz theorem}).

\section{Composite systems in topos quantum theory}\label{chap:composite topos}
In this section, we consider composite systems in topos quantum theory. In probability theory, the composite of random variables $X$ and $Y$ is defined by its product $X \times Y$.	A natural generalization of this composition to topos quantum theory would be given by the product of locales.	If the valuations on locales $X$ and $Y$ respectively correspond to quantum states on the associated Hilbert spaces $\hilb[X]$ and $\hilb[Y]$, a proper composition $X \times Y$ is expected to lead a bijective correspondence between valuations on $X \times Y$ and quantum states on $\hilb[X] \otimes \hilb[Y]$.

Although there seem to be many ways to define a product locale, we here choose one and test if it describes the composite quantum system.	It turns out that the valuations on our product locale has a bijective correspondence between positive over pure tensor states rather than quantum states.
This has an implication on positivity in our definition of composite systems in topos quantum theory.

Of course it is possible to construct the topos for $\hilb[X] \otimes \hilb[Y]$ regarded as a single system, and construct the locale $X_{\hilb[X] \otimes \hilb[Y]}$ whose valuations have bijective correspondence between quantum states on $\hilb[X] \otimes \hilb[Y]$.	This method is not of our interest since it neither makes a composition in topos quantum theory or is motivated from the composition in probability theory.

\subsection{Composite systems}\label{sec:composite}
The Gelfand duality suggests that taking the product of spaces $X$ and $Y$ is equivalent to taking the coproduct of corresponding algebras $\cont (X,\complex)$ and $\cont (Y,\complex)$, since we have
\begin{eqnarray}
\cont (X \times Y,\complex) = \cont (X,\complex) \otimes \cont (Y,\complex),
\end{eqnarray}
where the tensor on the right hand side represents the coproduct.
We try to define composition by products of spectra and coproducts of algebras motivated from the definition of composition in classical probability theory.
The generalization is, however, not straightforward and we do not obtain the unique definition.
Thus the composition of classical systems is described by a product of spaces or, equivalently, a coproduct of algebras.

\subsubsection{Coproducts of algebras}
If $A_i$ $(i=1,...,n)$ are noncommutative C*-algebras describing independent systems $\hil_i$, we have associated toposes $[\com(A_i),\sets]$ and Bohrifications $\uA[i]$ each internal to $[\com(A_i),\sets]$.	An obstacle in forming a coproduct of algebras $\uA[i]$ is that the toposes in which they are defined are different.
In Ref.\,\cite{WoltersHalvorson2013}, two unital commutative C*-algebras $\uA[1]$ in $[\com(A_1),\sets]$ and $\uA[2]$ in $[\com(A_2),\sets]$ are mapped to topos $[\com(A_1) \times \com(A_2), \sets]$ by geometric morphisms and composed there.	We follow this technique and present it in the way applicable for coproducts of more than three algebras.

Let $\com(A_1) \times ... \times \com(A_n)$ be a product of posets $\{ \com(A_i) \}_{i=1,...,n}$.
There is a geometric morphism $f^\ast_i : [\com(A_i),\sets] \to \topAi$ defined by
\begin{eqnarray}
f^\ast_i F (C_1,...,C_n) = F(C_i), \qquad ((C_1,...,C_n) \in \cAi)
\end{eqnarray}
for each objects $F$ in $[\com(A_i),\sets]$, which is a functor from $\com(A_i)$ to $\sets$.
If $f^\ast_i$ is applied on internal C*-algebra $\uA[i]$, object $f^\ast_i \uA[i]$ is a unital commutative C*-algebra internal to $\topAi$ given by
\begin{equation}
f^\ast_i \uA[i] (C_1,...,C_n) = C_i.
\end{equation}

We define the C*-algebra of the composite system in topos quantum theory by the coproduct of $f^\ast_i \uA[i]$.
\begin{thm}\label{thm:description of coproduct}
	Denote the coproduct of $f^\ast_i \uA[i]$ ($i=1,...,n$) in $\ccstar_{\topAi}$ by $f^*_1 \uA[1] \otimes ... \otimes f^*_n \uA[n]$.	As an object of $\topAi$, $f^*_1 \uA[1] \otimes ... \otimes f^*_n \uA[n]$ is explicitly given by
	\begin{equation}\label{eq:tensor of independent algebras}
	f^*_1 \uA[1] \otimes ... \otimes f^*_n \uA[n] (C_1,...,C_n) = C_1 \otimes ... \otimes C_n,
	\end{equation}
	for $(C_1,...,C_n) \in \cAi$, where the tensor product on the right hand side is for commutative C*-algebras in $\sets$.
\end{thm}
\emph{Proof})
See Appx.\,\ref{sec:coproduct}.
\qed
It is already shown that Eq.\,(\ref{eq:tensor of independent algebras}) defines the coproduct when $n=2$ \cite{WoltersHalvorson2013}.	We straightforwardly reinforced their result to coproducts of finitely many C*-algebras.

In summary, from the pairs of toposes and the Bohrification algebras in them $([\com(A_i),\sets], \uA[i])$ ($i=1,...,n$), we first make topos $[\prod_{i=1}^n \com(A_i),\sets]$. Its internal commutative C$^*$-algebra $f^*_1 \uA[1] \otimes ... \otimes f^*_n \uA[n]$ is obtained by taking the coproduct of $f^*_i \uA[i]$ ($i=1,...,n$) in the cateogry of commutative C$^*$-algebras internal to $[\prod_{i=1}^n \com(A_i),\sets]$.	The explicit description of $f^*_1 \uA[1] \otimes ... \otimes f^*_n \uA[n]$ is presented in Eq.\,(\ref{eq:tensor of independent algebras}).

\subsubsection{Gelfand duality and products of spectra}
Let $\underline{\Sigma}_{f^*_i \uA[i]}$ be the spectrum for $f^*_i \uA[i]$ obtained by the Gelfand duality in topos $[\prod_{i=1}^n \com(A_i),\sets]$.	These spectra are related to the spectrum for the coproduct $f^*_1 \uA[1] \otimes ... \otimes f^*_n \uA[n]$ by
\begin{eqnarray}
\underline{\Sigma}_{f^*_1 \uA[1] \otimes ... \otimes f^*_n \uA[n]} \cong \underline{\Sigma}_{f^*_1 \uA[1]} \times ... \times \underline{\Sigma}_{f^*_n \uA[n]},
\label{eq:Gelfand duality}
\end{eqnarray}
since the contravariant equivalence $\Sigma_{(-)}$ changes coproducts in $\mathbf{cCstar}_{[\prod_{i=1}^n \com(A_i),\sets]}$ into producs in $\mathbf{KCRegLoc}_{[\prod_{i=1}^n \com(A_i),\sets]}$.

There is a different way to take the product of Gelfand spactra, which gives an object not necessarily equal to $\underline{\Sigma}_{f^*_1 \uA[1]} \times ... \times \underline{\Sigma}_{f^*_n \uA[n]}$.	If the geometric morphisms $f^*_i$ are applied to spectra $\gA[i]$, the resulting objects $f^*_i \gA[i]$ are compact regular locales internal to $[\prod_{i=1}^n \com(A_i),\sets]$.
This is because the theory of Gelfand spectrum is geometric \cite{BanaschewskiMulvey2006}, and the Gelfand spectra is preserved by the geometric morphisms.
The compact regular locale $f^*_i \gA[i]$ is not necessarily equal to $\underline{\Sigma}_{f^*_i \uA[i]}$.	In other words, the following diagram does not necessarily commute:
\begin{equation}\label{eq:geometric morphism and Gelfand functor}
	\begin{aligned}\xymatrix{\mathbf{cCstar}_{[\com(A_i),\sets]} \ar[d]^{f^*_i} \ar[r]^\Sigma & \mathbf{KCRegLoc}_{[\com(A_i),\sets]} \ar[d]^{f^*_i} \\
		\mathbf{cCstar}_{[\prod_{i=1}^n \com(A_i),\sets]} \ar[r]^\Sigma & \mathbf{KCRegLoc}_{[\prod_{i=1}^n \com(A_i),\sets]}.
	}\end{aligned}
\end{equation}

Therefore their product $f^*_1 \underline{\Sigma}_{\uA[1]} \times ... \times f^*_n \underline{\Sigma}_{\uA[n]}$ taken in $\mathbf{KCRegLoc}_{[\prod_{i=1}^n \com(A_i),\sets]}$ is not necessarily equal to $\underline{\Sigma}_{f^*_1 \uA[1]} \times ... \times \underline{\Sigma}_{f^*_n \uA[n]}$.

Although $f^*_1 \underline{\Sigma}_{\uA[1]} \times ... \times f^*_n \underline{\Sigma}_{\uA[n]}$ may differ from $\underline{\Sigma}_{f^*_1 \uA[1]} \times ... \times \underline{\Sigma}_{f^*_n \uA[n]}$, it is another definition of composite systems in the Bohrification approach generalizing the composition of classical systems.	The fact that there is no unique way to extend the definition of composition to the Bohrification approach (and in fact, to any topos quantum theory) stems from the different toposes for marginal systems.	In classical measure theory, any marginal systems are objects in the unified topos $\sets$.	In the topos quantum theory, marginal systems are equipped with their own toposes.	When taking the coproduct of algebras or product of spectra, we need to send them to a unified topos by properly chosen maps such as geometric morphisms.	Then the ambiguity arises from the non-commutativity of the diagram (\ref{eq:geometric morphism and Gelfand functor}).

The remainder of this thesis only concerns the composite systems defined by the coproduct $f^*_1 \uA[1] \otimes ... \otimes f^*_n \uA[n]$ and product $\underline{\Sigma}_{f^*_1 \uA[1]} \times ... \times \underline{\Sigma}_{f^*_n \uA[n]}$.	To simplify notations, we denote them respectively by $\uAi$ and $\gAi$. The reason to choose these definitions for the composition is that we do not have a simple description of spectra $f^*_1 \underline{\Sigma}_{\uA[1]} \times ... \times f^*_n \underline{\Sigma}_{\uA[n]}$, while we have Eq.\,(\ref{eq:tensor of independent algebras}) for $\uAi$.

\begin{rem}
	The product on the right hand side of Eq.\,(\ref{eq:Gelfand duality}) does not change even if taken in the category of locales $\loc_{[\prod_{i=1}^n \com(A_i),\sets]}$ (not necessarily compact regular).	This is because the product of compact regular locales (such as $\underline{\Sigma}_{f^*_i \uA[i]}$) in the category of locales is again compact and regular \cite{Johnstone1981:tychonoff,Johnstone1986:stone}, and because compact regular locales are automatically completely regular in $[\prod_{i=1}^n \com(A_i),\sets]$.	As already mentioned in \cite{HeunenLandsmanSpitters2009}, internal compact regular locales are completely regular if the topos in question satisfies the axiom of dependent choice \cite{Johnstone1986:stone}, and every presheaf topos (including $[\prod_{i=1}^n \com(A_i),\sets]$) satisfies it \cite{FourmanScedrov1982:dependentchoice}.	We have to replace the underlying category for taking products of locales to $\loc_{[\prod_{i=1}^n \com(A_i),\sets]}$, when constructing Markov chains in Sec.\,\ref{chap:Markov}.
\end{rem}

\subsection{States on composite systems}\label{sec:jointstates}
The integrals over $\uA$ ($A = \B(\hil)$) correspond bijectively to density operators on $\hil$ \cite{HeunenLandsmanSpitters2009}.	This is an important evidence that the Bohrification approach describes quantum theory.	This bijective correspondence is not generalized straightforwardly to composite systems.	In the following, we show that integrals on composite systems corresponds to positive over pure tensor states.

\subsubsection{Integrals over coproducts of algebras}
Internal integrals over $\uA$ in $[\com(A),\sets]$ are natural transformations $I:\uA_{sa} \to \R$ whose components $I_C:\uA(C)_{sa} \to \R$ are all integrals in $\sets$ \cite{HeunenLandsmanSpitters2009}.	We first extend this fact to integrals over $\uAi$.
\begin{thm}[Integrals over $\uAi$]\label{thm:internal integrals}
	An integral over $\uAi$ is a family $\{ I_{(C_1,...,C_n)}:(C_1 \otimes ... \otimes C_n)_{sa} \rightarrow \R \}_{(C_1,...,C_n)\in \cAi}$ such that
	\begin{enumerate}
		\item each $I_{(C_1,...,C_n)}:(C_1 \otimes ... \otimes C_n)_{sa}  \rightarrow \R$ is an integral in $\sets$, and
		\item if $(C_1,...,C_n) \leq (C'_1 ,...,C'_n)$, then $I_{(C'_1,...,C'_n)}(a) = I_{(C_1,...,C_n)}(a)$ for all $a \in (C_1 \otimes ... \otimes C_n)_{sa}$.
	\end{enumerate}
\end{thm}
\emph{proof})
	See Appx.\,\ref{sec:internal integrals}.
\qed

Integrals over $\uAi$ correspond not to quantum states but to positive over pure tensor states.
\begin{dfn}[positive over pure tensor states]\label{dfn:POPT}
	Let $\hilb[i]$ $(i=1,...,n)$ be Hilbert spaces. A bounded linear operator $\omega$ on $\Hi$ is said to be a positive over pure tensor state if $\tr [\omega] = 1$ and
	\begin{eqnarray*}
	\tr[ (P_1 \otimes ... \otimes P_N) \omega ] \geq 0,
	\end{eqnarray*}
	is satisfied for any set of positive operators $P_i$ on $\hilb[i]$.
\end{dfn}
The set of all positive over pure tensor states on $\Hi$ is denoted by $\mW(\Hi)$.
\begin{thm}\label{thm:bijection integral POPT}
	There exists a bijective correspondence between positive over pure tensor states on $\Hi$ and integrals over $\uAi$, if the dimension of the Hilbert spaces $\hil_i$ are all at least three.	The integral $\{ I^\omega_{(C_1,...,C_n)}:(C_1 \otimes ... \otimes C_n)_{sa} \to \R \}_{(C_1,...,C_n)\in \cAi}$ corresponding to a positive over pure tensor state $\omega$ is defined by
	\begin{eqnarray}
	I^\omega_{(C_1,...,C_n)} (a) = \tr \omega a,
	\label{eq:from POPT to integral}
	\end{eqnarray}
	for all $a \in (C_1 \otimes ... \otimes C_n)_{sa}$ and $(C_1,...,C_n) \in \cAi$.
\end{thm}
\emph{proof})
	See Appx.\,\ref{sec:integrals and positive over pure tensor states}.
\qed

The gap between density operators and integrals over composite systems is understandable from the observation that $\uAi$ does not include contexts for entangled measurements.	Consider a context $C_A \otimes C_B$ in the bipartite $\uA[1] \otimes \uA[2]$ with $A_1 \cong \B(\complex_2)$ and $A_2 \cong \B(\complex_2)$ for example.	Marginal commutative C$^*$-algebras $C_A$ and $C_B$ are generated by certain $1$-dimensional projectors in $\complex_2$ as
\begin{eqnarray*}
C_A &=& \{ c_1 \ketbra[a_1] +c_2 \ketbra[a_2] ~|~ c_i \in \complex \},\\
C_B &=& \{ c_1 \ketbra[b_1] +c_2 \ketbra[b_2] ~|~ c_i \in \complex \}.
\end{eqnarray*}
Then any operator $O$ in $C_A \otimes C_B$ is decomposed uniquely to
\begin{eqnarray*}
\nonumber O && = c_{11} \ketbra[a_1] \otimes \ketbra[b_1] + c_{12} \ketbra[a_1] \otimes \ketbra[b_2] \\
&&+ c_{21} \ketbra[a_2] \otimes \ketbra[b_1] + c_{22} \ketbra[a_2] \otimes \ketbra[b_2].
\end{eqnarray*}
If we require $O$ to be positive semi-definite, all the coefficients $c_{ij}$ are non-negative.	As a consequence, all the positive operators included in $C_A \otimes C_B$ are separable.	Integrals over $\uAi$ are required to be positive by the condition (\ref{eq:positivity of integral}), but only for separable operators just like positive over pure tensor states.

\begin{rem}
	Although Thm.\,\ref{thm:bijection integral POPT} deals with the integrals over $\uAi$ internal to topos $\topAi$, the same bijective correspondence exists between positive over pure tensor states in $\mW (\hil_1 \otimes ... \otimes \hil_m)$ and integrals over $\uA[1] \otimes ... \otimes \uA[m]$ internal to $\topAi$ for any $m \leq n$.	This is because integrals over $\uA[1] \otimes ... \otimes \uA[m] : (C_1,...,C_n) \mapsto C_1 \otimes ... \otimes C_m$ in $\topAi$ ($m \leq n$), and integrals over $\uA[1] \otimes ... \otimes \uA[m] : (C_1,...,C_m) \mapsto C_1 \otimes ... \otimes C_m$ in $[\prod_{i=1}^m \com(A_i),\sets]$ are equivalent.
\end{rem}

\subsubsection{Riesz theorem and valuations on products of spectra}\label{ssec:product spectra}
From the constructive Riesz theorem, we have a locale isomorphism
\begin{eqnarray}
\val(\underline{\Sigma}_{\uAi}) \cong \integ(\uAi).
\label{eq:Riesz theorem}
\end{eqnarray}
Isomorphisms~(\ref{eq:Gelfand duality}) and (\ref{eq:Riesz theorem}) together imply
\begin{eqnarray}
\val(\gAi) \cong \integ(\uAi).
\label{eq:composite isomorphism}
\end{eqnarray}
In words, the locales of integrals over the coproduct algebra and of valuations on the corresponding product spectra are isomorphic.
The following corollary of Thm.\,\ref{thm:bijection integral POPT} is lead by this observation.
\begin{cor}\label{cor:POPT and valuations}
	There exists a bijective correspondence between positive over pure tensor states on $\Hi$ and valuations on $\gAi$, if the dimension of the Hilbert spaces $\hil_i$ are all at least three.
\end{cor}
The isomorphism (\ref{eq:composite isomorphism}) provides a way to analyze states on composite systems from two viewpoints.	In this section, we have used integrals over algebras, since they provide clearer expressions on the values they assign for observables.	In the next section, we use valuations on spectra to define Markov chains and to analyze their property.

\begin{rem}
	Although we have obtained a bijective correspondence between positive over pure tensor states and valuations (Cor.\,\ref{cor:POPT and valuations}), we do not know the explicit description of the valuation themselves.	This is partly because we did not obtain the explicit description of the product Gelfand spectrum, and also because it is difficult to interpret the bijective correspondence between integrals and valuations explicitly in out topos.	In the next section, rather than trying to calculate the Gelfand spectrum explicitly, we employ a categorical analysis for Markov chains so that we can keep the abstract argument from Cor.\,\ref{cor:POPT and valuations}.
\end{rem}

\section{Generalization of classical Markov chains by monads}\label{sec:markov chains for monads}
Random variables $X_1,~X_2,...,~X_n$ constitute of a Markov chain if the joint distribution $p$ on $X_1 \times X_2 \times ... \times X_n$ decomposes to
	\begin{equation}\label{eq:transition matrices for Markov chain}
	p(x_1,x_2,...,x_n) = f_{X_n|X_{n-1}}(x_{n-1})(x_n)...  f_{X_2|X_1}(x_1)(x_2) p_{X_1}(x_1),
	\end{equation}
	where $p_{X_1}$ is a probability distribution on $X_1$, and $f_{X_{i}|X_{i-1}}$ are transition matrices from $X_{i-1}$ to $X_i$ \cite{CoverThomas2006,Rivas2014:markov}.
	Expression (\ref{eq:transition matrices for Markov chain}) of Markov chains reveals a method to extend Markov chains to arbitrary lengths by transition matrices.
	In this section, we define Markov chains for general commutative monads by generalizing this expression, and study their properties without relying on the specific monads in use.
	We find a condition shared by the distribution monad and the valuation monad, that leads the triviality of our Markov chain in topos quantum theory later in Sec.\,\ref{sec:valuation}.

Our motivation to consider monads comes from the observation that the distribution monad \cite{Jacobs2011:distributionmonad} describes both probability distributions and the transition matrices.
The functor part $\dist:\sets \to \sets$ of the distribution monad can be identified with
\begin{eqnarray*}
\mathrm{system} \rightarrow \mathrm{states},
\end{eqnarray*}
assignment.
We formulate notions of systems, composite systems, states and joint states required for defining Markov chains in terms of monads and their Kleisli categories in Sec.\,\ref{sec:probability:monad:system states}.
Based on these notions, we define Markov chains for strong monads over cartesian categories in Sec.\,\ref{ssec:markov chains for strong moands}.	We also analyze the conditions on monads so that the defined Markov chains have similar properties to classical ones.

\subsection{Notions of systems and states}\label{sec:probability:monad:system states}
We first define ``systems'' as objects of a cartesian category $\category$ with product $\times:\category \times \category \to \category$ and the terminal object $1$.
Then the functor part $\func:\category \to \category$ of a monad $\monad$ on $\category$ is assumed to assign state spaces for systems.	States on $X$ are morphisms from the terminal object $1 \to \func X$ (equivalently, Kleisli morphisms $1 \to_\kl X$).

Composition of systems $X$ and $Y$ should be given by the product $X \times Y$, so that joint states are morphisms $1 \to \func (X \times Y)$.
We need canonical morphisms $\func (X \times Y) \to \func X$ and $\func(X \times Y) \to \func Y$ to define marginal states (like partial trace in quantum theory).	For the distribution monad, these canonical morphisms are defined by $\dist \pi_X$ and $\dist \pi_Y$, where $\pi_X: X \times Y \to X$ and $\pi_Y:X \times Y \to Y$ are projections for the cartesian product.	In the following we assume that the canonical morphisms for marginals are given by projectors $\func \pi$ for cartesian categories.

We further require the monad to be commutative with the unique Fubini map $\fub_{X,Y}:\func X \times \func Y \to \func (X \times Y)$ so that product states are well defined.
Since the Fubini map $\fub_{X,Y}:\dist X \times \dist Y \to \dist (X \times Y)$ of the distribution monad represents the construction of product states in $\dist (X \times Y)$ from two marginal states in $\dist X$ and $\dist Y$, we straightforwardly extend it to define product states on $\func (X \times Y)$ as those written by
\begin{equation}
\fub_{X,Y} \circ \langle p_X , p_Y \rangle,
\end{equation}
with pairs of marginal states $p_X : 1 \to_\kl X$ and $p_Y : 1 \to_\kl Y$, where $\langle p_X , p_Y \rangle$ represents the product of morphisms.
If the monad $\monad$ were strong but not commutative, there would be two candidates for product states $\dst \circ \langle p_x , p_Y \rangle$ and $\dst' \circ \langle p_x , p_Y \rangle$ which are not necessarily equal, where $\dst_{X,Y}:\func X \times \dist Y \to \func (X \times Y)$ and $\dst'_{X,Y}:\func X \times \func Y \to \func (X \times Y)$ are Fubini maps constructed from the strength and the costrength of the strong monad.

\subsection{Markov chains for commutative monads over cartesian categories}\label{ssec:markov chains for strong moands}
\begin{figure}[h]
	\centering
	\includegraphics[width=0.5\textwidth]{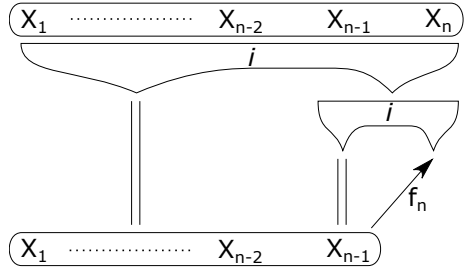}
	\caption{A configuration drawing of the process to extend Markov chains.	Two equality represent the unit morphism $\eta$ on the corresponding systems, and $f_n:X_{n-1} \to_\kl X_n$ represents the Kleisli morphism to extend the Markov chain.	Boxes labelled by $\fub$ denote the Fubini maps.}
	\label{fig:extension}
\end{figure}

Now we define Markov chains for strong monads on cartesian categories, by straightforwardly generalizing the classical ones.
We denote the Kleisli morphism $f:X \to \func Y$ of a monad $\monad$ by $f:X \to_\kl Y$ and the Kleilsi composition by $\odot$ to distinguish it from the usual composition $\circ$.
\begin{dfn}\label{def:Markov chain}
	Let $\category$ be a cartesian category with the terminal object $1$, and $\monad$ be a commutative monad with a Fubini map $\fub$.
	A joint state $m:1 \to \func (X_1 \times ... \times X_n)$ is said to be a Markov chain if there is a set of morphisms $\{ f_i :X_{i-1} \to_\kl X_i \}_{i=1,...,n}$ (with $X_0 := 1$) such that $m$ is equal to
	\begin{eqnarray}
	\ext_n \odot \ext_{n-1} \odot .... \odot \ext_3 \odot (\fub_{X_2,X_{1}} \circ \langle f_2,\eta_{X_1} \rangle) \odot f_1,
	\end{eqnarray}
	where
	\begin{eqnarray}
		\ext_i&:& X_{i-1} \times ... \times X_1 \to_\kl X_i \times X_{i-1} \times ... \times X_1\\
		\ext_i &:=& \fub_{X_i \times X_{i-1}, X_{i-2} \times ... \times X_1} \circ (\fub_{X_i,X_{i-1}} \circ \langle f_i,\eta_{X_i-1} \rangle \times \eta_{X_{i-2} \times ... \times X_1}).
		\label{eq:extension morphism}
	\end{eqnarray}

\end{dfn}
The morphisms $\ext_i$ ``extend'' Markov chains into larger Markov chains through the Kleisli morphism $f_i$ (see Fig.\,\ref{fig:extension} for a schematic representation).

For the case of classical probability theory, this definition of Markov chains for distribution monad on $\sets$ coincides with those of conventional definition presented by Eq.\,(\ref{eq:transition matrices for Markov chain}).	The transition matrices are generalized to Kleisli morphisms $f_i$.

While we have defined general Markov chains so that they coincide with the classical ones (\ref{eq:transition matrices for Markov chain}) if applied for probability distributions, the general ones may behave differently to the classical one.	Here we consider several properties we expect for general Markov chains to have, and derive sufficient conditions on the monad to define these desirable Markov chains.	Markov chains in topos quantum theory have these desirable properties since valuation monad is known to satisfy these sufficient conditions.

\paragraph{Stability of product states}
Let $p:1\to_\kl X \times Y$ be a bipartite state and $\ext:X \times Y \to_\kl X \times Y \times Z$ be defined by Eq.\,(\ref{eq:definition of ext}).	If $p$ is a product state then it would be natural to expect that $\ext \odot p$ is also a product state with regard to the bi-partition $X$ and $Y \times Z$. Furthermore the marginal state on $X$ should not be changed under the extension, since the extension map $\ext$ should be a local transformation on $Y$.	In words, product states are expected to be stable under the extension.	Product probability distributions are stable under the action of transition matrices.\footnote{Although trivial, any tripartite distribution defined by $p_{XYZ}(x,y,z) := f(y)(z)p_Y(y) p_X(x)$ with distributions $p_X$, $p_Y$ and transition matrix $f$ is again a product with respect to the bi-partition $X$-$YZ$.}

\begin{figure}[h]
	\centering
	\includegraphics[width=0.6\textwidth]{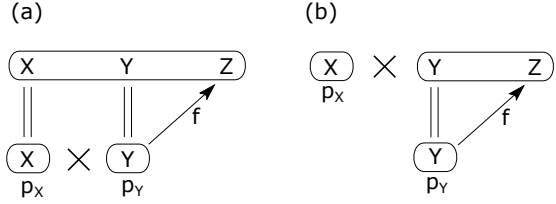}
	\caption{A schematic representation of Eq.\,(\ref{eq:extension preserves products}).	Equalities represent the unit morphism $\eta$ on the corresponding systems, $f$ is a Kleisli morphism. Symbols $\times$ are for product states.	The left and right hand sides of Eq.\,(\ref{eq:extension preserves products}) are expressed by (a) and (b), respectively.}
	\label{fig:extension3}
\end{figure}

The following lemma guarantees the stability of product states for any commutative monad.
\begin{lem}\label{lem:extending product states}
	Let $\monad$ be a commutative monad $\monad$ over cartesian category $\category$.	For all product state $i_{X,Y} \circ \langle p_X,~p_Y \rangle :1 \to_\kl X \times Y$ and all Kleisli morphisms $f:X \to_\kl X'$ and $g:Y \to_\kl Y'$, we have
	\begin{eqnarray}\label{eq:extendinf product states}
	(\fub_{X' , Y'} \circ (f \times g)) \odot (\fub_{X,Y} \circ \langle p_X,~p_Y \rangle) = \fub_{X' , Y'} \circ \langle f \odot p_X, g \odot p_Y \rangle.
	\end{eqnarray}
\end{lem}
\emph{proof})
	This lemma straightforwardly follows from the monoidal structure of $\fub$.	See Appx.\,\ref{sec:proof:lem:product} for an explicit proof.
\qed
This lemma states that local transformations preserves product states if the monad is commutative.	If applied to the extension $\ext$, Lem.\,\ref{lem:extending product states} suggests
\begin{eqnarray}
\ext \odot \langle p_X, p_Y \rangle = i_{X,Y \times Z} \circ \langle p_X, (\fub_{Y,Z} \circ \langle \eta_Y, f \rangle) \odot p_Y \rangle,
\label{eq:extension preserves products}
\end{eqnarray}
where the right hand side represents a product state as desired.	See Fig.\,\ref{fig:extension3} for a schematic representation of Eq.\,(\ref{eq:extension preserves products}).

\paragraph{Locality of extension}
A classical Markov chain has a property that the $n+1$-th random variable $X_{n+1}$ depends on $X_{n-1},~X_{n-2},...$ only through $X_n$ \cite{CoverThomas2006}.
If we require this property for our Markov chain, we expect that the marginal state
\begin{eqnarray}
\label{eq:created marginal}	&& \func \Pi_{Y \times Z} \circ (\ext \odot p), \\
&& \left( \ext := \fub_{X, Y \times Z} \circ (\eta_X \times \fub_{Y,Z} \circ \langle \eta_Y, f \rangle) \right)
\label{eq:definition of ext}
\end{eqnarray} 
with Kleisli morphisms $f:Y \to_\kl Z$ and $p:1 \to_\kl X \times Y$, depends on $p$ only through $p$'s marginal on $Y$.

The following lemma suggests an explicit construction of the marginal state given by (\ref{eq:created marginal}) from $\func \Pi_Y \circ p$.
\begin{lem}\label{lem:constructing marginals}
	Let $\ext:X \times Y \to_\kl X \times Y \times Z$ be defined as above by Eq.\,(\ref{eq:definition of ext}).
	The marginal state given by (\ref{eq:created marginal}) satisfies
	\begin{eqnarray}
	\nonumber \func \Pi_{Y \times Z} \circ (\ext \odot p) = (\fub_{Y,Z} \circ \langle \eta_Y, f \rangle) \odot (\func \Pi_Y \circ p).\\
	\label{eq:constructing marginals}
	\end{eqnarray}
	for any state $p:1 \to_\kl X \times Y$.
\end{lem}
\emph{proof})
	See Appx.\,\ref{sec:proof:lem:constructing marginals}.
\qed

\begin{figure}[h]
	\centering
	\includegraphics[width=0.6\textwidth]{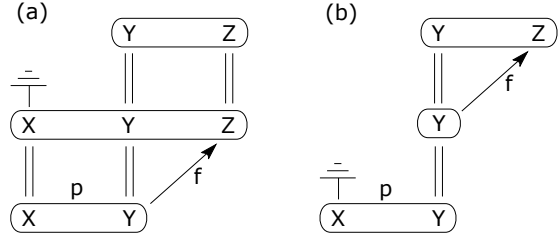}
	\caption{A schematic representation of Eq.\,(\ref{eq:constructing marginals}).	Equalities represent the unit morphism $\eta$ on the corresponding systems and $f$ is a Kleisli morphism.	The systems with ground symbols are discarded to take the marginal states on the other parts.	The left and right hand sides of Eq.\,(\ref{eq:constructing marginals}) are expressed by (a) and (b), respectively.}
	\label{fig:extension2}
\end{figure}

The right-hand-side of Eq.\,(\ref{eq:constructing marginals}) reveals that $\func \Pi_{Y \times Z} \circ (\ext \odot p)$ depends on $p$ only through its marginal on $Y$.\footnote{For the distribution monad, Eq.\,(\ref{eq:constructing marginals}) comes down to an almost trivial equation
	$\sum_{x \in X} p(x,y) f(y)(z) = f(y)(z) \sum_{x \in X} p(x,y)$.}
See Fig.\,\ref{fig:extension2} for a schematic representation of Eq.\,(\ref{eq:constructing marginals}).

\begin{figure}[h]
	\centering
	\includegraphics[width=0.3\textwidth]{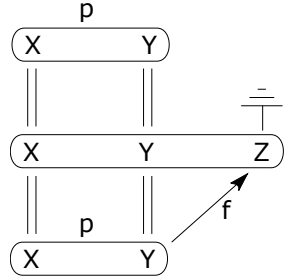}
	\caption{A schematic representation of Eq.\,(\ref{eq:marginal after extension2}).	Equalities represent the unit morphism $\eta$ on the corresponding systems, $f$ is a Kleisli morphism, and the marginal onto systems without the ground symbol is taken.}
	\label{fig:extension1}
\end{figure}

\paragraph{Preservation of original states}
While the morphism $\ext_i$ presented by Eq.\,(\ref{eq:extension morphism}) extends Markov chains, it might change the original state after extension.
More precisely, we are not sure if
\begin{eqnarray}
&& \func \Pi_{X \times Y} \circ ( \ext \odot p) = p
\label{eq:marginal after extension2} \\
&& \left( \ext := \fub_{X, Y \times Z} \circ (\eta_X \times \fub_{Y,Z} \circ \langle \eta_Y, f \rangle) \right)
\end{eqnarray}
holds for any Kleisli morphisms $f:Y \to_\kl Z$ and $p:1 \to_\kl X \times Y$, and any set of systems $\{X,~Y,~Z \}$.	See Fig.\,\ref{fig:extension1} for a schematic representation of the desired property (\ref{eq:marginal after extension2}).

We already know that this is satisfied for the distribution monad\footnote{If $p:X \times Y \to [0,1]$ is a joint distribution and $[f(y)(z)]_{y,z}$ is a transition matrix from $Y$ to $Z$, Eq.\,(\ref{eq:marginal after extension2}) reduces to $\sum_{z \in Z} p(x,y)f(y)(z) = p(x,y)$,	which is obvious from $\sum_{z \in Z} f(y)(z) = 1$.}.

We derive a sufficient condition on monad $\monad$ to satisfy Eq.\,(\ref{eq:marginal after extension2}) for any Markov chains.
\begin{lem}\label{lem:upper triangle}
	Equation (\ref{eq:marginal after extension2}) holds if $\func 1 \cong 1$.
\end{lem}
\emph{proof})
	See Appx.\,\ref{sec:proof:lem:upper triangle}.
\qed

In summary, the stability of product states under extension (\ref{eq:extension preserves products}) is guaranteed and the marginal of extended states coincides with the extension of marginal states (namely, Eq.\,(\ref{eq:constructing marginals}) holds) in any commutative monad over cartesian categories. Extensions preserve original states (namely, Eq.\,(\ref{eq:marginal after extension2}) holds) if $\func 1 \cong 1$.	We have obtained these results without relying on explicit descriptions of commutative monads and Kleisli morphisms therein. These results find their uses in the next section, where we do not have explicit description of Kleisli morphisms of the valuation monad.

\begin{rem}
	The distribution monad has
	\begin{eqnarray}\label{eq:normalization and affine}
	\dist 1 = \{ p:\{ \ast \} \to [0,1] ~|~p(\ast)=1 \} \cong 1.
	\end{eqnarray}
	There are many examples of commutative monads with $\func 1 \cong 1$ other than the distribution monad: Giry monad on measurable spaces \cite{Giry1982}, Radon monad on compact Hausdorff spaces \cite{FurberJacobs2015}, and the valuation monad on locales.	These examples are all somehow related to probability weight functions, where $\dist 1 \cong 1$ states the normalization of weight functions just as Eq.\,(\ref{eq:normalization and affine}) does.
\end{rem}
\begin{rem}
	There are commutative monads which do not have $\func 1 \cong 1$ but still validate Eq.\,(\ref{eq:marginal after extension2}).	An example which may be related to quantum physics is the Fock space monad $\mathcal{F}$ presented in Ref.\,\cite{Blute1994}.
\end{rem}
\begin{rem}\label{rem:classical extension problem}
	The extension of the bipartite distribution $\ext \odot p$ represents a solution of the ``extension problem'' for classical probability distributions.	The extension problem asks for necessary and sufficient conditions for a pair of bipartite states $p_{XY}$ on $X \times Y$ and $p_{YZ}$ on $Y \times Z$ to be expressed as marginal states of a tripartite state $p_{XYZ}$ (called an extension of $p_{XY}$ and $p_{YZ}$).	It is known that if $p_{XY}$ and $p_{YZ}$ are probability distributions, their extension probability distribution $p_{XYZ}$ exists if and only if they coincides on the common marginal $Y$ \cite{CarlenLebowitzLieb2013}.	An extension can be constructed by
	\begin{eqnarray}
	p_{XYZ}(x,y,z) := p_{Z|Y}(y)(z) p_{XY}(x,y),
	\end{eqnarray}
	where $p_{Z|Y}$ is the conditional distribution $p_{Z|Y}(y)(z) := p_{YZ}(y,z)/p_Y(y)$ regarded as a transition matrix.	This extension is equal to $\ext \odot p$ if we substitute $p_{XY}$ to $p$, and $p_{Z|Y}$ to $f$ inside $\ext$.
\end{rem}

\section{Transformations and Markov chains for topos quantum theory}\label{chap:Markov}
We have defined a composition of marginal systems for topos quantum theory, and analyzed general states on composite systems in Sec.\,\ref{chap:composite topos}.
This section is devoted to transformations on the valuations.
We analyze Markov chains in topos quantum theory for this purpose.

Kleisli morphisms of the \emph{valuation monad} are used in the place of transition matrices, and thus these morphisms are regarded as state transformations in topos quantum theory.
The Markov chains in topos quantum theory are expected to reflect properties of these morphisms.
While we are motivated to see the correspondence between these Kleisli morphisms and, for example, completely positive maps in usual quantum theory, our analysis on Markov chains rather indicates a triviality of these morphisms.

We first give an introduction to the valuation monad, and deduce an observation on the complete positivity of the Kleisli morphisms of the valuation monad from the bijective correspondence between multipartite valuations and positive over pure tensor states in Sec.\,\ref{sec:kleisli morphisms of valuation monad}.
Markov chains in topos quantum theory is considered in Sec.\,\ref{sec:valuation}.
We analyze monogamy properties of positive over pure tensor states independently to topos theory, and apply it to analyze these Markov chains.

\subsection{Kleisli morphisms of valuation monad}\label{sec:kleisli morphisms of valuation monad}
Valuation monad is a monad on category of locales $\loc$ presented by Vickers \cite{Vickers2011:monad}.
Vickers presents three types of valuation monads, representing unnormalized, subnormalized, and normalized valuations, among which we focus only on normalized valuations.
The functor part $\val:\loc \to \loc$ assigns the locale of valuations $\val X$ for a given locale $X$.	If $X$ is a compact, completely regular locale, $\val X$ coincides with the one presented in Ref.\,\cite{CoquandSpitters2009:integralvaluation} and reviewed in Sec.\,\ref{ssec:Gelfand and Riesz}.
The valuation monad is commutative and satisfies $\val 1 \cong 1$ much like the distribution monad. 

As we have reviewed in Sec.\,\ref{ssec:Gelfand and Riesz}, valuations on $X$ have bijective correspondence between $\Hom_\loc (1,X)$ where $1$ is the terminal object in $\loc$.
In other words, a valuation $v$ on locale $X$ is equivalent to a Kleisli morphism $v:1 \to_\kl X$ of the valuation monad, in the same way probability distribution $p$ on random variable $X$ is equivalent to a Kleisli morphism $p:1 \to_\kl X$ of the distribution monad.
The action of Kleisli morphism $f:X \to_\kl Y$ of the valuation monad on the valuation $v$ is defined by the Kleisli composition
\begin{equation}
f \odot v: 1 \xrightarrow{v} \val X \xrightarrow{\val f} \val^2 Y \xrightarrow{\mu_Y} \val Y
\end{equation}
in exactly the same way transition matrices acts on probability distributions.	Once a composition of transition matrices is written by Kleisli morphisms, its valuation version is immediately obtained by replacing the monad to valuations.

Let $\monad$ be a commutative monad over a cartesian category $\category$, $\fub$ be its Fubini map, and $f:X \to_\kl X'$ and $g:Y \to_\kl Y'$ be Kleisli morphisms.
The \emph{spatial composition} of $f$ and $g$ refers to a Kleisli morphism $f \otimes g : X \times Y \to_\kl X' \times Y'$ defined by
\begin{equation}
f \otimes g := \fub_{X',Y'} \circ (f \times g).
\end{equation}
For readers with familiarity on category theory, it suffices to say that Kleisli categories of commutative monads have monoidal structure $\otimes$ \cite{Goubaultlarrecq2008:monadictypes}.

All the Kleisli morphisms of a commutative monad is completely positive (CP) in the sense that arbitrary spatial composition of two Kleisli morphisms are allowed.
CP maps in quantum and classical probability theories are defined to send states to (possibly unnormalized) states, if it act in parallel with identity maps.
If we define the states on system $X \times Y$ to be Kleisli morphisms $v: 1 \to_\kl X \times Y$, the parallel action of the Kleisli morphism $f:X \to_\kl X'$ and the identity on $Y$ transforms it to
\begin{equation}
(f \otimes \eta_Y) \odot v,
\end{equation}
which is a state on $X' \times Y$ (note that unit $\eta_Y$ is the identity morphism on $Y$ in the Kleisli category).
Thus any Kleisli morphisms are valid state transformation even if they acts on marginal systems.

With regard to our composite system in topos quantum theory, valuations on $\gA[1] \times \gA[2]$ bijectively corresponds to positive over pure tensor states on $\hilb[1] \otimes \hilb[2]$, if $A_1 = \B(\hilb[1])$ and $A_2 = \B(\hilb[2])$.
Let $v: 1 \to_\kl \gA[1] \times \gA[2]$ be a valuation, and $f:\gA[1] \to_\kl \gA[1]$ be a Kleisli morphism of the valuation monad.
Then the Choi isomorphism between bipartite operators and linear maps \cite{Choi1975,Jiang2013} implies that a valuation
\begin{equation}
v' := (f \otimes \eta_{\gA[2]}) \odot v
\end{equation}
corresponds not necessarily to a quantum state, but to a positive over pure tensor state on $\hilb[1] \otimes \hilb[2]$.
Thus even if $f$ corresponds to a positive non-CP map in quantum theory and $v'$ does not corresponds to a quantum state, it is regarded to be CP in our definition of composite systems in topos quantum theory.
If there is another definition of composite systems that leads to a bijective correspondence between joint valuations and quantum states, Kleisli morphisms do not represent positive non-CP maps in quantum theory.

Nevertheless the above argument does not indicate the existence of Kleisli morphisms corresponding to positive maps.
It only denies the existence of Kleisli morphisms corresponding to non-positive maps.
More work is required to see which of the positive and CP maps in quantum theory, topos quantum theory leads when generalizing classical probability theory.

\subsection{Valuation monad for topos quantum theory}\label{sec:valuation}
This section applies results on Markov chains from Sec.\,\ref{sec:markov chains for monads} to the topos quantum systems of Sec.\,\ref{chap:composite topos}.
Category $\loc$ is cartesian \cite{Picado2004:locales}, and the valuation monad (of normalized valuations) is commutative and satisfies $\val 1 \cong 1$ \cite{Vickers2011:monad}, making this application possible.

Before going to Markov chains in topos quantum theory, we analyze properties of positive over pure tensor states in Sec.\,\ref{ssec:monogamy of POPT}.
We show that monogamy of multipartite quantum states, which is a characteristic property of quantum states that classical probability does not have, also exists for positive over pure tensor states.
From the monogamy of positive over pure tensor states, we deduce that every Markov chain of the valuation monad for topos quantum theory must be a product state in Sec.\,\ref{ssec:triviality of Markov chain}.
Thus we show a triviality of Markov chains.

\subsubsection{Monogamy of positive over pure tensor states}\label{ssec:monogamy of POPT}
In this subsection, we check that a monogamy property of quantum states also holds for positive over pure tensor states.	The results of this subsection is combined with those of Sec.\,\ref{ssec:markov chains for strong moands} to prove the triviality of Markov chains in topos quantum theory.

Monogamy in quantum theory means the following property of quantum correlations, namely, if a tripartite state has a strong correlation in one of its bipartite marginal system, then the marginal system does not have a strong correlation between the third.	This property of quantum correlations partly originates from non-trivial \emph{extendibility} conditions of marginal quantum states, which are considered in quantum marginal problems.	Among many extendibility situations, we focus on the simplest one by replacing quantum states into positive over pure tensor states.	We denote the set of positive over pure tensor states on composite system $\hilb[X] \otimes \hilb[Y] \otimes ...$ by $\mW(\hilb[X] \otimes \hilb[Y] \otimes ...)$.
\begin{dfn}
	A pair of bipartite positive over pure tensor states $\omega_{XY} \in \mW(\hilb[X] \otimes \hilb[Y])$ and $\omega_{YZ} \in \mW(\hilb[Y] \otimes \hilb[Z])$, is said to be \emph{extendible} if there exists a tripartite positive over pure tensor state $\omega_{XYZ} \in  \mW(\hilb[X] \otimes \hilb[Y] \otimes \hilb[Z])$ such that
	\begin{eqnarray}
	\tr_{\hilb[X]} [\omega_{XYZ}] = \omega_{YZ},\\
	\tr_{\hilb[Z]} [\omega_{XYZ}] = \omega_{XY}.
	\end{eqnarray} 
\end{dfn}
An obvious necessary condition for a pair $(\omega_{XY},\omega_{YZ})$ to be extendible is $\tr_{\hilb[X]}[\omega_{XY}] = \tr_{\hilb[Z]}[\omega_{YZ}]$, namely, they must coincide on the overlapping marginal.

The analogous concept of extendibility can be defined for classical probability distributions and quantum states as well \cite{CarlenLebowitzLieb2013}.	As we have remarked in Rem.\,\ref{rem:classical extension problem}, any pair of classical probability distributions with marginal states coinciding on the overlap is extendible.	This is not the case for quantum states (see e.g.~\cite{CarlenLebowitzLieb2013,TycVlach2015}), and positive over pure tensor states as we will see below.

The set of bipartite positive over pure tensor states is convex whose extremal points of which can be characterized~\cite{Hansen2013:extremalwitness}.
Examples of extremal bipartite positive over pure tensor states are pure quantum states, and partial transpositions of those.	
\begin{lem}\label{lem:nonextensibility POPT states 1}
	Let $\hilb[X],~\hilb[Y],~\hilb[Z]$ be Hilbert spaces, and $\omega \in \mW(\hilb[X] \otimes \hilb[Y] \otimes \hilb[Z])$ be a positive over pure tensor state.	If $\tr_{\hilb[Z]}(\omega)$ is an extremal bipartite positive over pure tensor state\footnote{Note that if $\omega \in \mW(\hilb[X] \otimes \hilb[Y] \otimes ...)$ is a positive over pure tensor state, its partial traces $\tr_{\hilb[i]}[\omega]$ are positive over pure tensor states.} on $\hilb[Y] \otimes \hilb[Z]$, then $\omega = \tr_{\hilb[Z]}(\omega) \otimes \tr_{\hilb[X] \otimes \hilb[Y]}(\omega)$. 
\end{lem}
\emph{proof})
	If $\{ O_i \in \B(\hilb[Z]) \}_{i=1,...,m}$ is a set of POVM measurement operators, then operators on $\hilb[X] \otimes \hilb[Y]$ defined by
	\begin{eqnarray}
	&& \omega|_i := \tr_{\hilb[Z]}[\omega (\iden_{\hilb[X] \otimes \hilb[Y]} \otimes O_i)] / p_i, \\
	&& (p_i := \trace[\omega (\iden_{\hilb[X] \otimes \hilb[Y]} \otimes O_i)] = \tr\left[  \tr_{\hilb[X] \otimes \hilb[Y]} [\omega  ] O_i \right] \geq 0)
	\end{eqnarray}
	are positive over pure tensor states on $H_1 \otimes H_2$, since
	\begin{eqnarray}
	\trace[\omega|_i ] = \frac{\trace[\omega (\iden_{\hilb[X] \otimes \hilb[Y]} \otimes O_i)]}{\trace[\omega (\iden_{\hilb[X] \otimes \hilb[Y]} \otimes O_i)]} = 1, \text{ and }\\
	\trace[\omega|_i (P_X \otimes P_Y)] = \tr_{\hilb[Z]}[\omega (P_X \otimes P_Y \otimes O_i)]/p_i \geq 0,	
	\end{eqnarray}
	hold for any positive operators $P_X$ and $P_Y$ on $\hilb[X]$ and $\hilb[Y]$.	The marginal state is represented by a convex combination of positive over pure tensor states
	\begin{eqnarray}
	\tr_{\hilb[Z]}[\omega] = \tr_{\hilb[Z]}[\omega (\iden_{\hilb[X] \otimes \hilb[Y]} \otimes \sum_{i=1}^m O_i)] = \sum_{i} p_i \omega|_i,
	\end{eqnarray}
	Since $\tr_{\hilb[Z]}[\omega]$ is assumed to be an extremal bipartite state, it follows that the states $\omega |_i$ are all equal to $\tr_{\hilb[Z]}[\omega]$.	Then for any set of positive operators $\{ P_X,~P_Y,~P_Z \}$ ($P_i \in  \hilb[i]$),
	\begin{eqnarray*}
	\trace[\omega (P_X \otimes P_Y \otimes P_Z)] &=& \tr \left[(P_X \otimes P_Y) \tr_{\hilb[Z]} [\omega (\iden_{\hilb[X] \otimes \hilb[Y]} \otimes P_Z)] \right] \\
	&=& \tr \left[(P_X \otimes P_Y) \tr_{\hilb[Z]} [\omega] \right] \times \tr \left[P_Z \tr_{{\hilb[X] \otimes \hilb[Y]}} [\omega] \right] \\
	&=& \tr \left[(P_X \otimes P_Y \otimes P_Z) \left( \tr_{\hilb[Z]} [\omega] \otimes \tr_{{\hilb[X] \otimes \hilb[Y]}} [\omega] \right)  \right]
	\end{eqnarray*}
	holds.	This implies $\omega =  \tr_{\hilb[Z]} [\omega] \otimes \tr_{{\hilb[X] \otimes \hilb[Y]}} [\omega]$ as claimed in the lemma. 
\qed
\begin{lem}\label{lem:nonextensibility POPT states 2}
	If either of a pair of nonproduct positive over pure tensor states $\omega_{XY} \in \mW(\hilb[X] \otimes \hilb[Y])$ or $\omega_{YZ} \in \mW(\hilb[Y] \otimes \hilb[Z])$ is extremal, the pair is not extendible.
\end{lem}
\emph{proof})
	Without loss of generality we assume that $\omega_{XY}$ is an extremal nonproduct positive over pure tensor state. If there is a tripartite state $\omega_{XYZ}$ with the property presented in the lemma, it follows from the previous lemma that $\omega_{XYZ}$ is expressed as $\omega_{XYZ} = \omega_{XY} \otimes \tr_{\hilb[X] \otimes \hilb[Y]}(\omega_{XYZ})$. Then $\tr_{\hilb[X]}[\omega_{XYZ}]$ is a product state, contradicting to the assumption that $\omega_{YZ}$ is a nonproduct state.
\qed

These lemmas can be interpreted as evidences of the monogamy of positive over pure tensor states.
A pair of bipartite positive over pure tensor states $\omega_{XY}$ and $\omega_{YZ}$ is not necessarily extendible even if they coincide on the overlap $\hilb[Y]$.
There are non-trivial extendibility conditions beyond the coincidence on overlaps, and this condition prohibits multipartite quantum states to have arbitrary strong correlations in all partitioning.
In particular, Lem.\,\ref{lem:nonextensibility POPT states 1} reveals that if a bipartite marginal state on a tripartite positive over pure tensor state is extremal, then the tripartite state must be a product of the bipartite marginal state and the rest.

We further analyze details of this monogamy property for showing a triviality of Markov chains in the next section.
\begin{thm}\label{lem:non-extendibility}
	For any non-product positive over pure tensor state $\omega_{XY}$ on $\hilb[X] \otimes \hilb[Y]$, there is a quantum state $\rho_{YZ}$ on $\hilb[Y] \otimes \hilb[Z]$ with $\dim \hilb[Z] = 3$ such that $\tr_{\hilb[X]}[\omega_{XY}] = \tr_{\hilb[Z]}[\rho_{YZ}]$ but the pair $(\omega_{XY},\rho_{YZ})$ is not extendible.
\end{thm}
\emph{proof})
	Denote the marginal state on $\hilb[Y]$ by $\rho_Y := \tr_{\hilb[X]} [\omega_{XY}]$. Define rank-$2$ projectors $\proj_{ij}$ ($i,j=1,...,\dim \hilb[Y]$, $i \neq j$) by
	\begin{eqnarray}
	\proj_{ij} = \ketbra[\psi_i] + \ketbra[\psi_j],
	\end{eqnarray}
	where $\{ \ket[\psi_i] \}_{i=1,...,\dim \hilb[Y]}$ is the basis diagonalizing $\rho_Y$ so that $\rho_Y = \sum_i p_i \ketbra[\psi_i]$.
	
	We first show that if $\omega_{XY}$ is not a product state, there exists a pair $i,j \in  \{1,...,\dim \hilb[Y] \}$ such that
	\begin{eqnarray}
	(\iden_{\hilb[X]} \otimes \proj_{ij}) \omega_{XY} (\iden_{\hilb[X]} \otimes \proj_{ij})
	\label{eq:two dimensional projections}
	\end{eqnarray}
	is not product.	If we assume the contrary,
	\begin{eqnarray}
	(\iden_{\hilb[X]} \otimes \proj_{ij}) \omega_{XY} (\iden_{\hilb[X]} \otimes \proj_{ij}) = O^{\hilb[X]}_{ij} \otimes O^{\hilb[Y]}_{ij}
	\end{eqnarray}
	holds for all pairs $i \neq j$.	Then for any one dimensional projector,
	\begin{eqnarray}
	\nonumber (\iden_{\hilb[X]} \otimes \ketbra[\psi_i]) \omega_{XY} (\iden_{\hilb[X]} \otimes \ketbra[\psi_i]) &=& (\iden_{\hilb[X]} \otimes \ketbra[\psi_i] \proj_{ij}) \omega_{XY} (\iden_{\hilb[X]} \otimes \proj_{ij} \ketbra[\psi_i]) \\
	&\propto& O^{\hilb[X]}_{ij} \otimes \ketbra[\psi_i].
	\end{eqnarray}	
	Since this holds for any $j$ not equal to $i$, we obtain $O^{\hilb[X]}_{ij} \propto O^{\hilb[X]}_{ij'}$ for any pair $j,j' (\neq i)$.	Since $O^{\hilb[X]}_{ij}$ is not changed under the permutation of $i$ and $j$, there exists an operator $O^{\hilb[X]}$ on $\hilb[X]$ such that
	\begin{eqnarray}
	O^{\hilb[X]} \propto O^{\hilb[X]}_{ij},
	\end{eqnarray}
	for any $i \neq j$.	This implies
	\begin{eqnarray*}
	\omega_{XY} &=& \sum_{i \neq j} (\iden_{\hilb[X]} \otimes \proj_{ij}) \omega_{XY} (\iden_{\hilb[X]} \otimes \proj_{ij}) - \sum_k (\iden_{\hilb[X]} \otimes \ketbra[\psi_k]) \omega_{XY} (\iden_{\hilb[X]} \otimes \ketbra[\psi_k])\\
	&\propto& O^{\hilb[X]} \otimes O^{\hilb[Y]},
	\end{eqnarray*}
	with an operator $O^{\hilb[Y]}$ on $\hilb[Y]$, which contradicts to the assumption that $\omega_{XY}$ is not product.
	
	Without loss of generality, we assume that the case of $i=1,j=2$ gives a non-product operator by Eq.\,(\ref{eq:two dimensional projections}).	Define $\ket[\phi]_{23}$ as an unnormalized purification of $p_1 \ketbra[\psi_1] + p_2 \ketbra[\psi_2]$
	\begin{eqnarray}
	\ket[\phi]_{YZ} := \sqrt{p_1} \ket[\psi_1] \otimes \ket[0]_Z + \sqrt{p_2} \ket[\psi_2] \otimes \ket[1]_Z,
	\end{eqnarray}
	and define $\rho_{YZ}$ by
	\begin{eqnarray}
	\rho_{YZ} := \ketbra[\phi]_{YZ} + \sum_{i\geq 3} p_i \ketbra[\psi_i] \otimes \ketbra[2]_Z,
	\end{eqnarray}
	so that the marginal state $\tr_{\hilb[Z]} [\rho_{YZ}]$ on $\hilb[Y]$ is equal to $\rho_Y$.	
	
	$\rho_{YZ}$ is not extendible with $\omega_{XY}$. Otherwise there exists a tripartite positive over pure tensor state $\omega_{XYZ}$ whose marginals are $\omega_{XY}$ and $\rho_{YZ}$.	The restriction of $\omega_{XYZ}$
	\begin{eqnarray}
	(\iden_{\hilb[X]} \otimes \proj_{12} \otimes \iden_{\hilb[Z]}) \omega_{XYZ} 	(\iden_{\hilb[X]} \otimes \proj_{12} \otimes \iden_{\hilb[Z]}),
	\end{eqnarray}
	must have marginals $(\iden_{\hilb[X]} \otimes \proj_{12}) \omega_{XY} (\iden_{\hilb[X]} \otimes \proj_{12})$ and $\ketbra[\phi]_{YZ}$.	This contradicts to Lem.\,\ref{lem:nonextensibility POPT states 2} since both of them are non-product and $\ketbra[\phi]_{YZ}$ is (proportional to) a pure state, which is an extremal point of $\mW(\hilb[Y] \otimes \hilb[Z])$.
\qed

\subsubsection{Triviality of Markov chains}\label{ssec:triviality of Markov chain}
We denote the Fubini map of valuation monad by $i_{X,Y}:\val X \times \val Y \to \val (X \times Y)$.	Similarly to the Fubini map for distribution monad, $i$ represents inclusion of product valuations to the space (locale, in precise) of valuations on composite system \cite{Vickers2011:monad}.	Product valuations on $X \times Y$ are those defined by
\begin{eqnarray}
i \circ \langle v_X , v_Y \rangle:1 \to_\kl X \times Y
\end{eqnarray}
with valuations $v_X :1 \to_\kl X$ and $v_Y: 1 \to_\kl Y$ on local systems $X$ and $Y$, respectively.

Now consider the valuation monad on the category of locales internal to the topos $[\prod_{i=1}^n \com(A_i),\sets]$, where $A_i = \B(\hilb[i])$ are non-commutative algebras of operators on finite dimensional Hilbert spaces.	Valuations on $\prod_{i \in I} \underline{\Sigma}_{A_i}$, where $I$ is a subset of $\{ 1,...,n \}$, correspond to positive over pure tensor states on $\bigotimes_{i \in I} \hilb[i]$ by Thm.\,\ref{thm:bijection integral POPT}.	Product valuations correspond to product quantum states.	Taking marginals of a valuation is equivalent to taking partial traces of the corresponding positive over pure tensor  state.

\begin{thm}\label{thm:triviality of Markov chains}
	Let $\hilb[i]$ ($i=1,...,n$) be finite dimensional Hilbert spaces all with dimensions at least $3$. Markov chains of valuations on $\gAi$ defined by Def.\,\ref{def:Markov chain}, are equivalent to product states by the bijective correspondence (\ref{eq:from POPT to integral}). 
\end{thm}
\emph{proof})
	Let $f_i:\gA[i-1] \to_\kl \gA[i]$ be Kleisli morphisms where $\gA[0] := 1$ for the terminal object $1$ of $\loc_{\topAi}$.	A sequence of Markov chains $v_i:1 \to_\kl \gA[1] \times ... \times \gA[i]$ are constructed by
	\begin{eqnarray}
	v_i := \ext_i \odot \ext_{i-1} \odot ... \odot \ext_3 \odot (i_{\gA[1],\gA[2]} \circ \langle \eta_{\gA[1]}, f_2 \rangle) \odot f_1,
	\label{eq:Markov chains of valuations}
	\end{eqnarray}
	where
	\begin{eqnarray}
	\ext_i := i_{\gA[1] \times ... \times \gA[i-2],\gA[i-1] \times \gA[i]} \circ (i_{\gA[i-1],\gA[i]} \circ \langle f_i,\eta_{\gA[i-1]} \rangle \times \eta_{\gA[1] \times ... \times \gA[i-2]}).
	\end{eqnarray}
	We inductively show that the Markov chains $v_i$ presented in Eq.\,(\ref{eq:Markov chains of valuations}) are product valuations and hence corresponds to product quantum states.
	
	We first deduce a contradiction by assuming the top one of this sequence
	\begin{eqnarray}
	v_2 = (i_{\gA[1],\gA[2]} \circ \langle \eta_{\gA[1]}, f_2 \rangle) \odot f_1
	\end{eqnarray}
	corresponds to non-product positive over pure tensor states.	We denote the quantum state corresponding to $v_2$ by $\rho_2$. If $\rho_2$ is not a product state, then $v_2$'s marginal $\val \Pi_{\gA[1]} \circ v_2$ corresponds to a mixed quantum state $\tr_{\hilb[2]}[\rho_2]$.	Lemma\,\ref{lem:non-extendibility} implies that there exists a valuation $v'_2$ on system $\gA[1] \times \gA[3] \times ... \times \gA[n]$ corresponding to a quantum state $\rho'_2$ whose marginal on $\hilb[1]$ is equivalent to $\tr_{\hilb[2]}[\rho_2]$, but $\rho_2$ and $\rho'_2$ are not marginal states of a single tripartite state.	Now consider a valuation $v$ on $\gAi$ defined by
	\begin{eqnarray}
	v:= \left( i_{\gA[1] \times \gA[2], \gA[3] \times ... \times \gA[n]} \circ (i_{\gA[1],\gA[2]} \circ \langle \eta_{\gA[1]}, f_2 \rangle) \right) \odot v'_2,
	\end{eqnarray}
	in words, we extend $v'_2$ by $f_2$.	Lemmas\,\ref{lem:upper triangle} and \ref{lem:constructing marginals} respectively imply
	\begin{eqnarray}
	\val \Pi_{\gA[1] \times \gA[2]} \circ v = v_2,\hspace{1cm} 
	\val \Pi_{\gA[1] \times \gA[3] \times ... \times \gA[n]} \circ v = v'_2.
	\end{eqnarray}
	These equations state that the positive over pure tensor  state corresponding to $v$ has marginal states $\rho_2$ and $\rho'_2$ overlapping at $\hilb[2]$, which contradicts to the assumption that $v_2$ and $v'_2$ corresponds to a non-extendible pair.	We have proven that $v_2$ is a product valuation.
	
	If $v_{i-1}$ is a product valuation, it is decomposed to
	\begin{eqnarray}
	v_{i-1} = i_{\gA[1] \times ... \times \gA[i-2], \gA[i-1]} \circ \langle w,w' \rangle,
	\end{eqnarray}
	where $w:1 \to_\kl \gA[1] \times ... \times \gA[i-2]$, and $w':1 \to_\kl \gA[i-1]$ are valuations on the marginal systems and $w$ is itself a product valuation.	Lemma\,\ref{lem:extending product states} suggests (cf. Eq.\,(\ref{eq:extension preserves products}))
	\begin{eqnarray}
	v_i &=& \ext_i \odot v_{i-1} \\
	&=& i_{\gA[1] \times ... \times \gA[i-2], \gA[i-1] \times \gA[i]} \circ \langle w, (i_{\gA[i-1],\gA[i]} \circ \langle f_i,\eta_{\gA[i-1]} \rangle) \odot w' \rangle.
	\end{eqnarray}
	Valuation $v_i$ is now written as a product of $w$ and
	\begin{eqnarray}
	w'' := (i_{\gA[i-1],\gA[i]} \circ \langle f_i,\eta_{\gA[i-1]} \rangle) \odot w'.
	\end{eqnarray}
	It suffices to check that $w''$ is a product state since $w$ is itself a product state.	This can be shown in the same way of showing that $v_2$ is a product state.
\qed

\begin{figure*}
	\centering
	\includegraphics[width=0.8\textwidth]{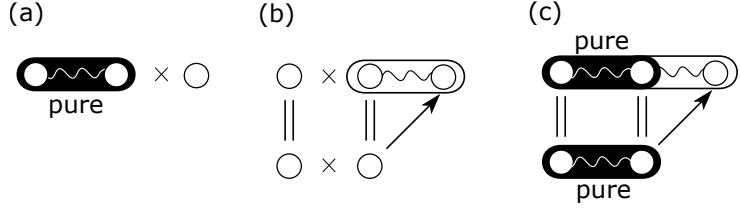}
	\caption{Three steps to show the triviality of our Markov chains.	(a) If a bipartite marginal of a tripartite state is correlated and pure, the third party must be disconnected to the other two.	(b) A Kleisli morphism which creates correlation.	(c) The same Kleisli morphism extends a bipartite correlated pure state to a state contradicting to (a).}
	\label{fig:monogamy}
\end{figure*}

The core of this proof can be abstractly summarized as follows (see Fig.\,\ref{fig:monogamy} alongside).	Extremal bipartite non-product states cannot be extended to tripartite positive over pure tensor states so that it is not product between the original two and the third party (Fig.\,\ref{fig:monogamy} (a)).	If there should exist a map $\ext$ which extends product states to non-product positive over pure tensor states (Fig.\,\ref{fig:monogamy} (b)), however, it would extend extremal bipartite non-product state to create the forbidden tripartite state (Fig.\,\ref{fig:monogamy} (c)).	The proof shows a fundamental incompatibility between our Markov chains and monogamy of states.	Once a bijective correspondence between states defined by any commutative monad with $\func 1 \cong 1$ and quantum states (or positive over pure tensor states) on composite Hilbert spaces is established, the monogamy of quantum states (or positive over pure tensor states) immediately indicates the triviality of Markov chains.	

The fundamental incompatibility between our Markov chains and the monogamy property stems from the cartesianness of the underlying category.	Cartesianness of the underlying category enables us to construct morphisms
\begin{eqnarray}
i \circ \langle f,g \rangle
\end{eqnarray}
by combining \emph{arbitrary} Kleisli morphisms $f$ and $g$.	That means, every pair of Kleisli morphisms is compatible in the sense these two can be combined together to form a $1$-to-$2$ map whose marginals coincides with the original pair.	This is in contrast to the TPCP maps where a pair of two maps with the same domain is not necessarily compatible \cite{JohnsonViola2015:channelextension}.

Theorem\,\ref{thm:triviality of Markov chains} also implies the non-existence of Kleisli morphisms $f:\gA[i-1] \to_\kl \gA[i]$ that create non-product states between $\gA[i-1]$ and $\gA[i]$.
	Typical morphisms which create product states between $\gA[i-1]$ and $\gA[i]$ are
\begin{eqnarray}
f_{v}: \gA[i-1] \xrightarrow{!_{\gA[i-1]}} 1 \xrightarrow{v} \val \gA[i], 
\label{eq:factoring through 1}
\end{eqnarray}
with the unique morphism $!_{\gA[i-1]}$ to the terminal object and any valuation $v:1 \to_\kl \gA[i]$ on $\gA[i]$.
In words, $f_v$ outputs a fixed state $v$ no matter what the input is.	If $f_v$ is used for extension, it just adds state $v$ as the last member of the Markov chain.
Although we do not have a proof, we conjecture that the only Kleisli morphisms from $\gA[i-1]$ to $\gA[i]$ are those given by Eq.\,(\ref{eq:factoring through 1}).

\begin{rem}
	There is another generalization of classical Markov chains to quantum theory, which exhibits classical correlations unlike our Markov chains for topos quantum theory.	A tripartite quantum state $\rho_{XYZ} \in \state(\hilb[X] \otimes \hilb[Y] \otimes \hilb[Z])$	is a \emph{short quantum Markov chain} \cite{Hayden2004:ssa} if there is a TPCP map $\E^{\rho_{XYZ}}_{Y \to YZ}:\B(\hilb[Y]) \to \B(\hilb[Y] \otimes \hilb[Z])$ such that
	\begin{equation*}
	\rho_{XYZ} = \id_{\hilb[X]} \otimes \E^{\rho_{XYZ}}_{Y \to YZ} \left( \tr_{\hilb[Z]} [\rho_{XYZ}] \right) .
	\end{equation*}
	There are short quantum Markov chains which are not product in any partitioning.	The trick to avoid the triviality forced by the monogamy is the dependency of $\E^{\rho_{XYZ}}_{Y \to YZ}$ on state $\rho_{XYZ}$.	The map $\id_{\hilb[X]} \otimes \E^{\rho_{XYZ}}_{Y \to YZ}$ extending chains preserves the marginal state $\tr_{\hilb[Z]} [\rho_{XYZ}]$ of $\rho_{XYZ}$, but it may change other input states from $\state(\hilb[X] \otimes \hilb[Y])$.	If $\E^{\rho_{XYZ}}_{Y \to YZ}$ must satisfy $\tr_{\hilb[Z]}[\id_{\hilb[X]} \otimes \E^{\rho_{XYZ}}_{Y \to YZ}(\rho_{XY})] = \rho_{XY}$ for any state $\rho_{XY}$ in $\state(\hilb[X] \otimes \hilb[Y])$, $\E^{\rho_{XYZ}}_{Y \to YZ}$ separates to $\E_{Y} \otimes \Gamma_{\rho_Z}$ with a TPCP map $\E_{Y}:\B(\hilb[Y]) \to \B(\hilb[Y])$ and $\Gamma_{\rho_Z}:\B(\complex) \to \B(\hilb[Z])$ to prepare state $\rho_Z$, and the resulting short quantum Markov chain is product, much like our Markov chains for topos quantum theory.
\end{rem}

\section{Concluding remarks}\label{chap:conclusion}
We defined composite systems in topos quantum theory, by generalizing products of random variables representing the composite systems in classical probability theory.	From toposes and their internal commutative C*-algebras describing marginal quantum systems, a unifying topos is first constructed, and the coproduct of C*-algebras is taken in the unifying topos, as done for bipartite systems in Ref.\,\cite{WoltersHalvorson2013}.	Taking coproducts of C*-algebras is equivalent to taking product of corresponding locales.	The joint valuations on product locales and, equivalently, the integrals over coproduct algebras have bijective correspondence between positive over pure tensor states instead of quantum states.

The gap between joint valuations and quantum states arises since the joint valuations are not required to be positive on entangled positive operators, while quantum states are.	This is because our coproducts of commutative C*-algebras lack the commutative C*-subalgebra for entangled observables.	It is open if there exists another definition of composite systems leading a bijective correspondence between joint valuations and quantum states.	Commutative C*-algebras for such a composition would include sufficient commutative subalgebras for entangled observables.

Our analysis reveals that there is no unique way to generalize composite systems of classical probability theory to topos quantum theory.	This arbitrariness arises in the first place from the use of different toposes for describing marginal quantum systems.	All random variables in classical probability theory are objects in unique topos $\sets$, while the marginal locales of topos quantum theory may exist in different toposes.

The Kleisli morphisms of the valuation monad are regarded to be state transformations in topos quantum theory, and we have investigated its behaviour on composite systems.	The commutativity of valuation monad alone implies that the Kleisli morphisms are CP in the sense that arbitrary two Kleisli morphisms acting in parallel generate valid Kleisli morphisms.	This does not imply, however, that positive non-CP quantum maps are excluded from the Kleisli morphisms of the valuation monad.	Since joint valuations corresponding to positive over pure tensor states are regarded as states in our definition of composite systems, actions of positive non-CP maps on a part of entangled states produces valid states in topos quantum theory.	The definition of complete positivity for topos quantum theory highly depends on how to define composite systems.

We generalized classical Markov chains directly to topos quantum theory, by replacing distribution monad (used in classical probability theory) to valuation monad (used in topos quantum theory).	The Markov chains are recursively defined by extending short Markov chains by Kleisli morphisms of valuation monad.	We have shown a fundamental incompatibility between our Markov chains and the monogamy property of positive over pure tensor states, and demonstrated that our Markov chains correspond only to product valuations on composite systems.	Markov chains in topos quantum theory is more trivial than classical ones which can contain classical correlation.	This consequence reveals that there only exist state transformations between different marginal systems that do not create correlation in topos quantum theory.
Kleisli morphisms between different marginal systems seems to be too trivial to ask the correspondence between linear maps in quantum theory. 

This triviality of Markov chains has two origins: the use of (cartesian) product to describe composite systems, and $1 \cong \val 1$ for the valuation monad.	It is sometimes considered that the product may not be suitable for describing composition for quantum systems, for example because of the no-cloning theorem \cite{CoeckePaquette2011:practisingphysicist}.	The triviality of our Markov chains reinforces this observation, by showing an incompatibility between product and the monogamy existing in quantum states and positive over pure tensor states.	While we cannot define our Markov chains without the product, $1 \cong \val 1$ is not a crucial property.	It might be interesting to consider our Markov chains for different monad $\func$ and cartesian category other than the valuation monad on category of locales, such that $1 \ncong \func 1$.

\appendix

\section{Proofs for theorems}

\subsection{Coproducts of internal unital commutative C*-algebras}\label{sec:coproduct}
This appendix shows Thm.\,\ref{thm:description of coproduct} by generalizing the method presented in Ref.\,\cite{WoltersHalvorson2013}.	We shall prove the following Lemma which directly implies the theorem:
\begin{lem}\label{lem:description of coproduct}
	Let $P$ be a poset and functors $A_i:P \rightarrow \sets$ $(i=1,...,n)$ be unital commutative C*-algebras internal to the functor topos $[P, \sets]$.	Then the object $A_1 \otimes ... \otimes A_n: P \rightarrow \sets$ defined by
	\begin{eqnarray}
	A_1 \otimes ... \otimes A_n (x) = A_1 (x) \otimes ... \otimes A_n (x), \label{eq:tensor general objects}
	\end{eqnarray}
	for elements (objects) $x \in P$ and
	\begin{eqnarray}
	A_1 \otimes ... \otimes A_n (f) (a_1 \otimes ... \otimes a_n ) = A_1 (f)(a_1) \otimes ... \otimes  A_n (f)(a_n),	\label{eq:tensor general arrows}\\
	\nonumber (a_i \in A_i (x) ~(\forall i))
	\end{eqnarray}
	for partial orders (morphisms) $f:x \xrightarrow{\leq} y$, where the tensor products on the right hand sides of Eqs.\,(\ref{eq:tensor general objects}) and (\ref{eq:tensor general arrows}) are for unital commutative C*-algebras in $\sets$, is an internal commutative C*-algebra.	Furthermore it is the coproduct of $\{ A_i \}_{i=1,...,n}$ in the category of internal unital commutative C*-algebras.
\end{lem}

It is shown in Ref.\,\cite{SpittersVickersWolters2014} that an object $A$ in $[\T, \sets]$ is an internal unital commutative C*-algebra if and only if each component $A(x)$ is a unital commutative C*-algebra in $\sets$ and the arrow part $A(f): A(x) \rightarrow A(y)$ for any $f:x \rightarrow y$ in $\T$ is a unital *-homomorphism, where $\T$ is a small category.	This implies that the object $A_1 \otimes \cdots \otimes A_n$ is an internal unital commutative C*-algebra, since each component $A_1 \otimes ... \otimes A_n (X)$ is a unital commutative C*-algebra in $\sets$, and a tensor product of unital *-homomorphisms is again a unital *-homomorphism.	Note that poset $P$ is a small category.

Interpretation of theories for unital *-homomorphisms by the Kripke-Joyal semantics on $[P,\sets]$ (see e.g.~\cite{MacLaneMoerdijk1992} for details) reveals that a natural transformation $\alpha \colon A \to B$ between internal unital commutative C*-algebra objects $A$, $B$ in $[P,\sets]$ is an internal *-homomorphism if and only if all the components are *-homomorphisms of unital commutative C*-algebras in $\sets$.	For example, linearity
\[
\forall a,b \in A,  \alpha(a) + \alpha(b) = \alpha(a + b)
\]
holds 
\begin{eqnarray*}
\text{iff } && \bot \Vdash \forall a,b \in A \colon \alpha(a) + \alpha(b) = \alpha(a + b), \\
\text{iff } && \forall p \in P,~ \forall a \in A(p),~ p \Vdash \forall b \in A \colon \alpha(a) + \alpha(b) = \alpha(a + b), \\
\text{iff } && \forall p \in P,~ \forall a \in A(p),~\forall q \geq p,~ \forall b \in A(q),~ q \Vdash \alpha(a|_q) + \alpha(b) = \alpha(a|_q + b), \\
\text{iff } && \forall p \in P,~ \forall a \in A(p),~\forall q \geq p,~ \forall b \in A(q),~ \alpha_q (a|_q) + \alpha_q (b) = \alpha_q (a|_q + b).
\end{eqnarray*}
This statement is equivalent to the simpler statement
\begin{equation}
\forall p \in P,~ \forall a,b \in A(p),~ \alpha_p (a) + \alpha_p (b) = \alpha_p (a+b).
\label{eq:presheaf semantics1}
\end{equation}
Thus the linearity on $\alpha$ reduces to the linearity of each component.
The other axioms for *-homomorphisms are interpreted similarly, and reduce to component-wise axioms in $\sets$.

Define natural transformations $\alpha^i \colon A_i \to A_1 \otimes \cdots \otimes A_n$ as the candidate coproduct injections by setting
\begin{equation}
\alpha^i_p (a_i) = \iden_1 \otimes \cdots \otimes \iden_{i-1} \otimes a_i \otimes \iden_{i+1} \otimes \cdots \otimes \iden_n.
\label{eq:injection}
\end{equation}
These natural transformations are internal *-homomorphisms because each component is.

Now let $A$ be any internal unital commutative C*-algebra, with internal *-homomorphisms $\beta^i \colon  A_i  \to A$.	Consider morphisms $\gamma_p \colon A_1(p) \otimes \cdots \otimes A_n (p) \to A(p)$ defined for $p \in P$ by
\[
\gamma_p (a_1 \otimes \cdots \otimes a_n) = \beta^1_p (a_1) \beta^2_p (a_2) \cdots \beta^n_p (a_n).
\]
This is a natural transformation since for $f \colon p \to q$ we have
\begin{align*}
\gamma_q (A_1 \otimes \cdots \otimes A_n(f)(a_1 \otimes\cdots \otimes a_n) ) 
&= \gamma_q ( A_1(f)(a_1) \otimes \cdots \otimes A_n(f)(a_n) ) \\
&= \beta^1_y (A_1(f)(a_1)) \beta^2_q (A_2(f)(a_2)) \cdots \beta^n_q (A_n(f)(a_n)) \\
&= A(f)(\beta^1_p (a_1)) A(f)(\beta^2_p (a_2)) \cdots A(f)(\beta^n_p (a_n)) \\
&= A(f)(\beta^1_p (a_1) \beta^2_p (a_2) \cdots \beta^n_p (a_n)) \\
&= A(f)(\gamma_p (a_1 \otimes \cdots \otimes a_n)).
\end{align*}
Clearly $\gamma \circ \beta^i = \alpha^i$ holds, and since each component is unique, $\gamma$ is the unique mediating map satisfying this condition.

\subsection{Internal integrals}\label{sec:internal integrals}
We show that the internal integrals over $\uAi$ is given by Def.\,\ref{thm:internal integrals}, by interpreting the axioms for integrals by the Kripke-Joyal semantics.
The real number object $\uR$ in topos $[P,\sets]$ is the constant functor which assigns the real numbers in $\sets$ for each $p \in P$ (see e.g.~\cite{MacLaneMoerdijk1992}).

Note first that the unit of multiplication $\iden$ for $\uAi$ is given by the component wise identities $\iden_{C_1 \otimes ... \otimes C_n}$ ($(C_1,...,C_n) \in \cAi$), and the unit $1$ of the internal real number object $\uR$ is also given by the component wise unit.	This implies that the normalization condition $I(\iden)=1$ of the natural transformation $I:\uAi \to \uR$ holds if and only if $I_C(\iden_{C_1 \otimes ... \otimes C_n}) = 1$ for all $C_1 \otimes ... \otimes C_n \in \cAi$.	Linearity and positivity are shown in the following manner.

Linearity of $I:\uAi \to \uR$
\[
\forall a,b \in \uAi \colon  I(a) + I(b) = I(a + b)
\]
holds 
\begin{eqnarray*}
\text{iff } && \bot \Vdash \forall a,b \in \uAi,  I(a) + I(b) = I(a + b), \\
\text{iff } && \forall p \in \cAi,~ \forall a \in \uAi(p),~ p \\
&&\Vdash \forall b \in \uAi \colon I(a) + I(b) = I(a + b), \\
\text{iff } && \forall p \in \cAi,~ \forall a \in \uAi(p),~\forall q \geq p,\\
&&~ \forall b \in \uAi(q),~ q \Vdash I(a|_q) + I(b) = I(a|_q + b).
\end{eqnarray*}
This statement is equivalent to
\begin{eqnarray*}
&&\forall p \in \cAi,~ \forall a \in \uAi(p),~\forall q \geq p,\\
&&~ \forall b \in \uAi(q),~ I_q (a|_q) + I_q (b) = I_q (a|_q + b),
\end{eqnarray*}
which is further equivalent to a simpler statement
\begin{equation*}
\forall p \in \cAi,~ \forall a,b \in \uAi(p),~ I_p (a) + I_p (b) = I_p (a+b).
\end{equation*}
Thus the natural transformation $I:\uAi \to \uR$ is linear if and only if its components are all linear.

Positivity of $I:\uAi \to \uR$
\[
\forall a \in \uAi \colon  a \geq 0 \Rightarrow I(a) \geq 0
\]
holds 
\begin{eqnarray*}
\text{iff } && \bot \Vdash \forall a \in \uAi \colon  a \geq 0 \Rightarrow I(a) \geq 0, \\
\text{iff } && \forall p \in \cAi,~ \forall a \in \uAi(p),~ p \Vdash a \geq 0 \Rightarrow I(a) \geq 0, \\
\text{iff } && \forall p \in \cAi,~ \forall a \in \uAi(p),~\forall q \geq p,~\\
&& q \Vdash a|_q \geq 0 \text{ implies } q \Vdash I(a|_q) \geq 0.
\end{eqnarray*}
Under the component-wise definition of internal C*-algebras \cite{SpittersVickersWolters2014}, this statement is equivalent to
\begin{equation*}
\forall p \in \cAi,~ \forall a \in \uAi(p),~\forall q \geq p,~ a|_q \geq 0 \text{ implies } I_q(a|_q) \geq 0
\end{equation*}
which is further equivalent to a simpler statement
\begin{equation*}
\forall p \in \cAi,~ \forall a \in \uAi(p),~ a \geq 0 \text{ implies } I_p(a) \geq 0.
\end{equation*}
Thus the natural transformation $I:\uAi \to \uR$ is positive if and only if its components are all positive.

In summary, a natural transformation $I:\uAi \to \uR$ is an integral over $\uAi$ if and only if its components are all integrals in $\sets$.	The naturality of $I$ is equivalent to the lower condition in Def.\,\ref{thm:internal integrals}.

\subsection{Integrals and positive over pure tensor states}\label{sec:integrals and positive over pure tensor states}
In this section, we present the proof of Theorem~\ref{thm:bijection integral POPT} which asserts the bijective correspondence between integrals and positive over pure tensor states.	We first introduce the notion of \emph{unentangled frame functions}, and a generalization of Gleason's theorem called the \emph{unentangled Gleason's theorem}, which are required for the proof.

Functions which have values on unentangled bases of composite Hilbert spaces have been previously analyzed under the name of {\it unentangled frame functions}.	More precisely, they are defined as follows.
\begin{dfn}[unentangled frame function \cite{RudolphWright2000,Wallach2002}]
	Let $\mH_i$ $(i=1,...,N)$ be Hilbert spaces, and ${\rm Prod} ( \mH_1 ,..., \mH_N)$ be the set of all product unit vectors on $\mH_1 \otimes ...\otimes \mH_N$.	An unentangled frame function for $\mH_1 ,...,\mH_N $ is a function $f: {\rm Prod}(\mH_1,...,\mH_N) \rightarrow \R^+$ such that for some positive number $w$ (called the weight of $f$), $\sum_j f(\xi_j) = w$ holds whenever $\left\lbrace \xi_j \right\rbrace_j$ is an orthonormal basis of $\mH_1 \otimes... \otimes \mH_N$ with each $\xi_j \in {\rm Prod}(\mH_1,...,\mH_N)$.
\end{dfn}
We denote the set of unit-weight unentangled frame functions for $\mH_1 ,...,\mH_N$ by $\UFF$.
The notion of unentangled frame function is an extension of the frame function defined in Ref.\,\cite{Gleason1975}, where the correspondence between quantum states on a single system and frame functions on the system is shown.	For composite systems, the following theorem is known.
\begin{thm}\cite[Thm.\,1]{Wallach2002}\label{thm:UGT}
	Let $\mH_1 ,...,\mH_N $ be finite-dimensional Hilbert spaces of each dimension at least $3$. Let $f: {\rm Prod}(\mH_1,...,\mH_N) \rightarrow \R^+$ be an unentangled frame function. Then there exists a self-adjoint operator $\omega_f$ in $\mathcal{B}(\Hi)$ such that whenever $v_1 \otimes ... \otimes v_N $ is in ${\rm Prod}(\mH_1,...,\mH_N)$ and $p_i$ is the projection of $\mH_i$ onto the one-dimensional subspace generated by $v_i$,
	\begin{eqnarray}
	f(v_1 \otimes ... \otimes v_n) = \tr (p_1 \otimes ... \otimes p_n) \omega_f.
	\label{eq:from UFF to POPT}
	\end{eqnarray}
\end{thm}
The uniqueness of $\omega_f$ for a given $f$ in this Theorem is shown in Ref.\,\cite{RudolphWright2000}, although it is not explicitly mentioned in Ref.\,\cite{Wallach2002}.	

Theorem~\ref{thm:UGT} implies the bijective correspondence between positive over pure tensor states and unit-weight unentangled frame functions, and called ``unentangled Gleason's theorem''.	Under this correspondence, Theorem~\ref{thm:bijection integral POPT} is rewritten in terms of the unentangled frame functions as follows: an injective map from $\UFF$ to integrals defined by
\begin{eqnarray}
f \mapsto \{ I^f_{(C_1,...,C_N)}:(C_1 \otimes ... \otimes C_N)_{sa} \rightarrow \R \}_{(C_1,...,C_N)\in \cAi},\\
I^f_{(C_1,...,C_N)} (p_1 \otimes ... \otimes p_N) = f(v_1 \otimes ... \otimes v_n) \hspace{1cm} (\text{if $p_i \in C_i$ for all $i$}),
\label{eq:from UFF to integral}
\end{eqnarray}
where $p_i$ is the projector onto the one-dimensional subspace generated by $v_i$, is a bijection if the dimension of all the Hilbert spaces $\mH_i$ is at least 3.

We show that a map from integrals to $\UFF$ presented by
\begin{eqnarray}
\{ I_{(C_1,...,C_N)}:(C_1\otimes ,..., \otimes C_N)_{sa} \rightarrow \R \}_{(C_1,...,C_N)\in \cAi} \mapsto f^I,\\
f^I(v_1 \otimes ... \otimes v_n) = I_{(C_1,...,C_N)} (p_1 \otimes ... \otimes p_N)  \hspace{1cm} (\text{if $p_i \in C_i$ for all $i$}),
\label{eq:from integral to UFF}
\end{eqnarray} 
is well-defined and injective.	First observe for any integral $\{ I_{(C_1,...,C_N)} \}_{(C_1,...,C_N)\in \cAi}$, and for all product projectors $p_1 \otimes ... \otimes p_N \in (C_1\otimes ,..., \otimes C_N)_{sa} \cap (C'_1\otimes ,..., \otimes C'_N)_{sa}$, we have
\begin{eqnarray*}
I_{(C_1,...,C_N)} (p_1 \otimes ... \otimes p_N) = I_{(C_1 \cap C'_1,...,C_N \cap C'_N)} (p_1 \otimes ... \otimes p_N) = I_{(C'_1,...,C'_N)} (p_1 \otimes ... \otimes p_N),
\end{eqnarray*}
since $p_1 \otimes ... \otimes p_N \in (C_1\otimes ,..., \otimes C_N)_{sa} \cap (C'_1\otimes ,..., \otimes C'_N)_{sa}$ implies $p_1 \otimes ... \otimes p_N \in (C_1 \cap C'_1 \otimes ... \otimes C_N \cap C'_N)_{sa}$.	Thus the value of integration does not depend on the context, and the map presented by Eq.~(\ref{eq:from integral to UFF}) is well-defined.	Furthermore, the map is injective, since the context-wise linearity of integrals reveals that an integral is uniquely determined by its value on product projectors.

The map defined by Eq.~(\ref{eq:from integral to UFF}) is the inverse of the map defined by Eq.~(\ref{eq:from UFF to integral}).	This completes the proof of Theorem~\ref{thm:bijection integral POPT}.

\subsection{Proof of Lem.\,\ref{lem:extending product states}}\label{sec:proof:lem:product}
We prove Lem.\,\ref{lem:extending product states}.	If we represent the Kleisli composition in the left hand side of Eq.\,(\ref{eq:extendinf product states}) by the usual composition in $\category$, the left hand side is transformed to the right hand side via
\begin{eqnarray}
&& \mu_{X' \times Y'} \circ \func (\fub_{X' \times Y'} \circ (f \times g)) \circ \fub_{X,Y} \circ \langle p_X,~p_Y \rangle \\
&=& \mu_{X' \times Y'} \circ \func \fub_{X' \times Y'} \circ \func (f \times g) \circ \fub_{X,Y} \circ \langle p_X,~p_Y \rangle \\
&=& \mu_{X' \times Y'} \circ \func \fub_{X' \times Y'} \circ \fub_{\func X,\func Y} \circ (\func f \times \func g) \circ  \langle p_X,~p_Y \rangle, \\
&=& \fub_{X' \times Y'} \circ (\mu_{X'} \times \mu_{Y'}) \circ (\func f \times \func g) \circ \langle p_X,~p_Y \rangle \\
&=& \fub_{X' \times Y'} \circ \langle f \odot p_X, g \odot p_Y \rangle.
\end{eqnarray}
Here, the second equality comes from the naturality of $i$.	The third equality comes from the commutativity of
\[
\xymatrix{\func^2 X \otimes \func^2 Y \ar[r]^{\fub_{\func X, \func Y}} \ar[d]_{\mu_A \otimes \mu_B} & \func(\func X \otimes \func Y) \ar[r]^{\func \fub_{X,Y}} & \func^2 (A \otimes B) \ar[d]^{\mu_{X \otimes Y}} \\
	\func X \otimes \func Y \ar[rr]^{\fub_{X,Y}} && \func (X \otimes Y),
}
\]
for any pair of $X$ and $Y$, which is shown for any commutative monad on symmetric monoidal categories in Ref.\,\cite{Goubaultlarrecq2008:monadictypes}.

\subsection{Proof of Lem.\,\ref{lem:constructing marginals}}\label{sec:proof:lem:constructing marginals}
In this section, we prove Lem.\,\ref{lem:constructing marginals}.	The left hand side of Eq.\,(\ref{eq:constructing marginals}) is transformed into
\begin{eqnarray}
\func \pi_{Y \times Z} \circ \mu_{X \times Y \times Z} \circ \func \ext \circ p &=& \mu_{Y \times Z} \circ \func^2 \pi_{Y \times Z} \circ \func \ext \circ p \\ &=& \mu_{Y \times Z} \circ \func(\func \pi_{Y \times Z} \circ \ext) \circ p,
\end{eqnarray}
where the first equality comes from the naturality of $\mu$.	The right hand side of Eq.\,(\ref{eq:constructing marginals}) is transformed into
\begin{eqnarray}
\mu_{Y \times Z} \circ \func (\fub_{Y,Z} \circ \langle \eta_X,f\rangle ) \circ \func \pi_Y \circ p = 	\mu_{Y \times Z} \circ \func (\fub_{Y,Z} \circ \langle \eta_X,f\rangle \circ \pi_Y) \circ p.
\end{eqnarray}
Thus Eq.\,(\ref{eq:constructing marginals}) holds if
\begin{eqnarray}
\func \pi_{Y \times Z} \circ \ext = \fub_{Y,Z} \circ \langle \eta_X,f\rangle \circ \pi_Y.\label{eq:constructing marginals proof1}
\end{eqnarray}

Although the theorem assumes commutativity of the monad, for a moment it is convenient to distinguish two Fubini maps
\begin{eqnarray*}
	\dst_{X,Y}&:& \func X \otimes \func Y \xrightarrow{\cst_{X,\func Y}} \func(X \otimes \func Y) \xrightarrow{\func \str_{X,Y}} \func^2 (X \otimes Y) \xrightarrow{\mu_{X \otimes Y}} \func (X \otimes Y),\\
	\dst'_{X,Y}&:&\func X \otimes \func Y \xrightarrow{\str_{\func X,Y}} \func(\func X \otimes Y) \xrightarrow{\func \cst_{X,Y}} \func^2 (X \otimes Y) \xrightarrow{\mu_{X \otimes Y}} \func (X \otimes Y),
\end{eqnarray*}
which coincides to $\fub_{X,Y}$ on commutative monads.
Here $\str_{X,Y}:X \times \func Y \to \func(X \times Y)$ and $\cst_{X,Y}:\func X \times Y \to \func (X \times Y)$ represent the strength and the costrength of the monad (see e.g.~\cite{Moggi1991}).
Consider the following diagram:{\small
	\[
	\xymatrix@C+3ex{
		X \times Y \ar[rr]^{\id_X \times \fub_{Y,Z} \circ \langle \eta_Y,f\rangle} \ar[d]^{!_X \times \id_Y} && X \times \func (Y \times Z) \ar[r]^{\eta_X \times \id_{\func (Y \times Z)}} \ar[d]_{!_X \times \func \id_{Y \times Z}} \ar@/_2pc/[rr]^{\str_{X , Y \times Z}} & \func X \times \func (Y \times Z) \ar[r]^{\dst'_{X,Y \times Z}} & \func(X \times Y \times Z) \ar[d]_{\func (!_X \times \id_{Y \times Z})} \\
		1 \times Y \ar[rr]^{\id_1 \times \fub_{Y,Z} \circ \langle \eta_Y, f \rangle} \ar[d]^{\pi_Y} && 1 \times \func (Y \times Z) \ar[rr]^{\str_{1,Y \times Z}} \ar[drr]^{\pi_{\func(Y \times Z)}} && \func (1 \times Y \times Z) \ar[d]_{\func \pi_{Y \times Z}} \\
		Y \ar[rrrr]^{\fub_{Y,Z} \circ \langle \eta_Y,f \rangle} &&&& \func (Y \times Z).	
	}
	\]
}
The compositions of arrows from $X \times Y$ to $\func (X \times Y)$, going lower-left edges and upper-right edges represent the right and left hand sides of Eq.\,(\ref{eq:constructing marginals proof1}).	The left two squares commute by the definition of product, the lower right triangle by the definition of the strength, the right square by the naturality of strength, and the upper right triangle by Lem.\,\ref{lem:Kock says this is trivial} shown below.	Thus all the triangles and squares commute and we have shown Eq.\,(\ref{eq:constructing marginals proof1}) when $\fub$ is $\dst'$.

\begin{lem}\cite{Kock1971:bilinearity}\label{lem:Kock says this is trivial}
	Triangles
	\begin{equation}\label{dg:Kock says this is trivial}
	\xymatrix{\func X \times Y \ar[d]_{\id_X \times \eta_Y} \ar[r]^{\cst_{X,Y}} &\func (X \times Y)  & X \times \func Y \ar[d]_{\eta_X \times \id_Y} \ar[r]^{\str_{X,Y}} & \func (X \times Y)\\
		\func X \times \func Y \ar[ur]^{\dst_{X,Y}} & & \func X \times \func Y \ar[ur]^{\dst'_{X,Y}} & 
	}
	\end{equation}
	commute.
\end{lem}
\emph{Proof})
Although the commutativity of these diagrams is suggested in Ref.\,\cite{Kock1971:bilinearity}, we provide an explicit proof since we were unable to find it in the literature.	By decomposing $\dst$ according to its definition, the left triangle decomposes into
\[
\xymatrix{
	\func X \times Y \ar[r]^{\cst_{X,Y}} \ar[d]_{\id_X \times \eta_Y} & \func (X \times Y) \ar[d]_{\func(\id_X \times \eta_Y)} \ar[dr]^{\func \eta_{X \times Y}} \ar[r]^{\id_{\func(X \times Y)}} & \func (X \times Y) \\
	\func X \times \func Y \ar[r]^{\cst_{X,\func Y}} & \func (X \times \func Y) \ar[r]^{\func \str_{X,Y}} & \func^2 (X \times Y). \ar[u]_{\mu_{X \times Y}}
}\]
The left square commutes by the naturality of $\cst$, the lower triangle by the unit law of strength, and the upper triangle is the unit law of monad.
The commutativity of the right triangle can be shown by a symmetric argument.

\subsection{Proof of Lem.\,\ref{lem:upper triangle}}\label{sec:proof:lem:upper triangle}
In this section, we give a proof of Lem.\,\ref{lem:upper triangle}.	Equation\,(\ref{eq:marginal after extension2}) is rewritten into
\begin{eqnarray}
p = \func \pi_{X \times Y} \circ \mu_{X \times Y \times Z} \circ \func \ext \circ p &=& \mu_{X \times Y} \circ \func^2 \pi_{X \times Y} \circ \func \ext \circ p, \\
&=& \mu_{X \times Y} \circ \func (\func \pi_{X \times Y} \circ \ext) \circ p,
\end{eqnarray}
where the second equality comes from the naturality of $\mu$. This holds if $\func \pi_{X \times Y} \circ \ext = \eta_{X \times Y}$, in other words, if the following diagram commute:
\begin{equation}\label{dg:upper triangle proof1}
\begin{aligned}\xymatrix@C+2ex{&&& & \func (X \times Y)  \\
	X \times Y \ar@/^1pc/[urrrr]^{\eta_{X \times Y}} \ar[rrr]^{\eta_X \times \fub_{Y,Z} \circ \langle \eta_Y,f \rangle} &&& \func X \times \func(Y \times Z) \ar[r]^{\fub_{X, Y \times Z}}  & \func(X \times Y \times Z). \ar[u]_{\func \pi_{X \times Y}}
}\end{aligned}
\end{equation}

Although $\fub = \dst = \dst'$ holds for commutative monad, it is convenient to substitute $\fub =\dst$ for a moment.	Consider following decomposition of diagram (\ref{dg:upper triangle proof1}) into pieces
\begin{equation}\label{dg:upper triangle proof2}
\begin{aligned}\xymatrix@C+2ex{&&& \func X \times Y \ar[r]^{\cst_{X,Y}} \ar[d]^{\id_{\func X} \times \eta_Y} & \func (X \times Y)  \\
	&&& \func X \times \func Y \ar[ur]_{\dst_{X \times Y}} & \\
	X \times Y \ar@/^5pc/[uurrrr]^{\eta_{X \times Y}} \ar[rrr]^{\eta_X \times \fub_{Y,Z} \circ \langle \eta_Y,f \rangle} \ar[urrr]^{\eta_X \times \eta_Y} \ar[uurrr]^{\eta_X \times \id_Y} &&& \func X \times \func(Y \times Z) \ar[r]^{\dst_{X, Y \times Z}} \ar[u]_{\func \id_X \times \func \pi_Y} & \func(X \times Y \times Z). \ar[uu]_{\func \pi_{X \times Y}}
}\end{aligned}
\end{equation}
The triangle at the top represents the unit law for $\cst$. The centre left triangle is trivial.	The square commutes from the naturality of $\dst$, because $\func \pi_{X \times Y} = \func (\id_X \times Y)$.	The centre right triangle is in Lem.\,\ref{lem:Kock says this is trivial}.	The lower left triangle is the only one whose commutativity is not shown.

By the definition of product $\times$, the lower left triangle decomposes into two triangles
\[
\xymatrix@C+3ex{&&& \func Y &&& \func X \\
	Y \ar[rr]^{\langle \eta_Y, f \rangle} \ar@/^1pc/[urrr]^{\eta_Y} && \func Y \times \func Z \ar[r]^{\fub_{Y,Z}} & \func(Y \times Z), \ar[u]_{\func \pi_Y} & X \ar[rr]^{\eta_X} \ar@/^1pc/[urr]^{\eta_X} && \func X. \ar[u]_{\func \id_X}
}
\]
The right triangle clearly commutes. Thus diagram (\ref{dg:upper triangle proof1}) commutes if the left triangle commutes, which is guaranteed by Lem.\,\ref{lem:good Fubini maps} shown below.	This completes the proof.
\begin{lem}\label{lem:good Fubini maps}
	If $\func 1 \cong 1$ for a strong monad $\monad$ on a cartesian category, the following diagrams commute:
	\begin{equation}\label{dg:good Fubini maps}
	\begin{aligned}\xymatrix@R-2ex{& \func X \\
		\func X \times \func Y \ar[ur]^{\pi_{\func X}} \ar[r]_{\dst_{X,Y}} & \func(X \times Y) \ar[u]_{\func \pi_X},
	}\end{aligned} \hspace{1cm}
	\begin{aligned}\xymatrix@R-2ex{	\func X \times \func Y \ar[r]^{\dst'_{X,Y}} \ar[dr]_{\pi_{\func Y}} & \func(X \times Y) \ar[d]^{\func \pi_Y} \\
		& \func Y,
	}\end{aligned}
	\end{equation}
	for any objects $X$ and $Y$.
\end{lem}
\emph{Proof})
This lemma is already known for cartesian \emph{closed} categories \cite{Kock1971:bilinearity}.	We here extend this known result on cartesian closed categories into cartesian categories not necessary closed.	We prove the left triangle, writing $!:Y \to 1$. Note that projection $\pi_X:X \times Y \to X$ can be decomposed into $X \times Y \xrightarrow{\id_X \times !} X \times 1 \xrightarrow{\pi_X} X$. It follows from naturality of $\mathrm{dst}$ that
\begin{eqnarray*}
&& \func \pi_X \circ \mathrm{dst}_{X,Y} 
= \func (\pi_X \circ \id_X \times !) \circ \mathrm{dst}_{X,Y}  \\
&& = \func \pi_X \circ \mathrm{dst}_{X,1} \circ \func \id_X \times \func ! 
= \func \pi_X \circ \mathrm{dst}_{X,1} \circ \id_{\func X} \times \func !,
\end{eqnarray*}
so it suffices to show $\func \pi_X \circ \mathrm{dst}_{X,1} = \pi_{\func X}$.
But this follows from the proof of~\cite[Thm.~2.1]{Kock1971:bilinearity}; notice that while that result assumes cartesian closedness, only cartesianness is sufficient for our purpose.	The right triangle of (\ref{dg:good Fubini maps}) can be shown by a symmetric argument.

\subsection*{Acknowledgements}
I would like to thank Chris Heunen, Mio Murao and Akihito Soeda for helpful discussions and comments on this paper, and Jonathan Barrett, Raymond Lal, Matty Hoban and Marco T{\'u}lio Quintino for useful comments on positive over pure tensor states.

This work is supported by the Leading Graduate Course for Frontiers of Mathematical Sciences and Physics, the Ministry of Education, Culture, Sports, Science and Technology, Japan, and by Japan Society for the Promotion of Science, Grants-in-Aid for Scientific Research, Grant Number 15J11531.


\begin{thebibliography}{99}

\bibitem{Accardi1982:quantumstochastic}
Accardi, L., Frigerio, A., Lewis, J.T.: Quantum stochastic processes.
\newblock Publications of the Research Institute for Mathematical Sciences
\textbf{18}(1), 97--133 (1982)

\bibitem{BaeChuscinski2016}
Bae, J., Chru{\'s}ci{\'n}ski, D.: Operational characterization of divisibility
of dynamical maps.
\newblock Phys. Rev. Lett. \textbf{117}, 050,403 (2016)

\bibitem{BanaschewskiMulvey2006}
Banaschewski, B., Mulvey, C.J.: A globalisation of the gelfand duality theorem.
\newblock Ann. Pure Appl. Log. \textbf{137}(1 -- 3), 62 -- 103 (2006)

\bibitem{Barnum2010:popt}
Barnum, H., Beigi, S., Boixo, S., Elliott, M.B., Wehner, S.: Local quantum
measurement and no-signaling imply quantum correlations.
\newblock Phys. Rev. Lett. \textbf{104}, 140,401 (2010)

\bibitem{Blute1994}
Blute, R.F., Panangaden, P., Seely, R.A.G.: Holomorphic models of exponential
types in linear logic.
\newblock In: S.~Brookes, M.~Main, A.~Melton, M.~Mislove, D.~Schmidt (eds.)
Mathematical Foundations of Programming Semantics. MFPS 1993., \emph{Lecture
	Notes in Computer Science}, vol. 802, pp. 474--512. Springer, Berlin,
Heidelberg (1994)

\bibitem{BuscemiDatta2016}
Buscemi, F., Datta, N.: Equivalence between divisibility and monotonic decrease
of information in classical and quantum stochastic processes.
\newblock Phys. Rev. A \textbf{93}, 012,101 (2016)

\bibitem{CarlenLebowitzLieb2013}
Carlen, E.A., Lebowitz, J.L., Lieb, E.H.: On an extension problem for density
matrices.
\newblock J. Math. Phys. \textbf{54}(6), 062103 (2013)

\bibitem{Caspers2009:nlevel}
Caspers, M., Heunen, C., Landsman, N.P., Spitters, B.: Intuitionistic quantum
logic of an n-level system.
\newblock Found. Phys. \textbf{39}(7), 731--759 (2009)

\bibitem{Choi1975}
Choi, M.D.: Completely positive linear maps on complex matrices.
\newblock Linear Algebra Appl. \textbf{10}(3), 285 -- 290 (1975)

\bibitem{CoeckePaquette2011:practisingphysicist}
Coecke, B., Paquette, {\'E}.: Categories for the practising physicist.
\newblock In: B.~Coecke (ed.) New Structures for Physics, pp. 173--286.
Springer-Verlag Berlin Heidelberg (2011)

\bibitem{CoquandSpitters2009:integralvaluation}
Coquand, T., Spitters, B.: Integrals and valuations.
\newblock Log. Anal. \textbf{1}(3), 1--22 (2009)

\bibitem{CoverThomas2006}
Cover, T.M., Thomas, J.A.: Elements of Information Theory.
\newblock Wiley, New York (1991)

\bibitem{DoringIsham2011}
D{\"o}ring, A., Isham, C.: ``what is a thing?'': Topos theory in the
foundations of physics.
\newblock In: B.~Coecke (ed.) New Structures for Physics, pp. 753--937.
Springer-Verlag Berlin Heidelberg (2011)

\bibitem{DoringIsham2008I}
{D{\"o}ring}, A., Isham, C.J.: {A topos foundation for theories of physics}.
\newblock J. Math. Phys. \textbf{49}(5), 053515 (2008)

\bibitem{FerrieEmerson2009:framedspace}
Ferrie, C., Emerson, J.: Framed hilbert space: hanging the quasi-probability
pictures of quantum theory.
\newblock New J. Phys. \textbf{11}(6), 063,040 (2009)

\bibitem{FourmanScedrov1982:dependentchoice}
Fourman M.P., S.A.: The "world's simplest axiom of choice" fails.
\newblock Manuscripta Math. \textbf{38}, 325--332 (1982)

\bibitem{FurberJacobs2015}
Furber, R.W.J., Jacobs, B.P.F.: From {K}leisli categories to commutative
{C}*-algebras: probabilistic {G}elfand duality.
\newblock Log. Meth. Comput. Sci. \textbf{11}(2), 5 (2015)

\bibitem{Giry1982}
Giry, M.: A categorical approach to probability theory.
\newblock In: B.~Banaschewski (ed.) Categorical Aspects of Topology and
Analysis, \emph{Lecture Notes in Mathematics}, vol. 915, pp. 68--85.
Springer, Berlin, Heidelberg (1982)

\bibitem{Gleason1975}
Gleason, A.M.: Measures on the closed subspaces of a hilbert space.
\newblock In: C.~Hooker (ed.) The Logico-Algebraic Approach to Quantum
Mechanics, \emph{The University of Western Ontario Series in Philosophy os
	Science}, vol.~5a, pp. 123--133. Springer Netherlands (1975)

\bibitem{Goubaultlarrecq2008:monadictypes}
Goubault-Larrecq, J., Lasota, S., Nowak, D.: Logical relations for monadic
types.
\newblock Math. Struct. Comput. Sci. \textbf{18}(6), 1169--1217 (2008)

\bibitem{Hayden2004:ssa}
Hayden, P., Jozsa, R., Petz, D., Winter, A.: Structure of states which satisfy
strong subadditivity of quantum entropy with equality.
\newblock Commun. Math. Phys. \textbf{246}(2), 359--374 (2004)

\bibitem{Henry2015:geometricbohr}
{Henry}, S.: {A Geometric Bohr topos}.
\newblock arXiv:1502.01896  (2015)

\bibitem{HeunenLandsmanSpitters2009}
Heunen, C., Landsman, N.P., Spitters, B.: A topos for algebraic quantum theory.
\newblock Commun. Math. Phys. \textbf{291}(1), 63--110 (2009)

\bibitem{IshamButterfield1998}
Isham, C.J., Butterfield, J.: Topos perspective on the kochen-specker theorem:
I. quantum states as generalized valuations.
\newblock Int. J. Theor. Phys. \textbf{37}(11), 2669--2733 (1998)

\bibitem{Jacobs2011:distributionmonad}
Jacobs, B.: Probabilities, distribution monads, and convex categories.
\newblock Theor. Comput. Sci. \textbf{412}(28), 3323 -- 3336 (2011)

\bibitem{Jiang2013}
Jiang, M., Luo, S., Fu, S.: Channel-state duality.
\newblock Phys. Rev. A \textbf{87}, 022,310 (2013)

\bibitem{JohnsonViola2015:channelextension}
Johnson, P.D., Viola, L.: On state versus channel quantum extension problems:
exact results for {$U\;\otimes \;U\;\otimes \;U$} symmetry.
\newblock J. Phys. A: Math. Theor. \textbf{48}(3), 035,307 (2015)

\bibitem{Johnstone1981:tychonoff}
Johnstone, P.: Tychonoff's theorem without the axiom of choice.
\newblock Fund. Math. \textbf{113}(1), 21--35 (1981)

\bibitem{Johnstone1986:stone}
Johnstone, P.T.: Stone spaces, \emph{Cambridge Studies in Advanced
	Mathematics}, vol.~3.
\newblock Cambridge University Press, Cambridge (1986)

\bibitem{KochenSpecker1975}
Kochen, S., Specker, E.P.: The problem of hidden variables in quantum
mechanics.
\newblock In: C.A. Hooker (ed.) The Logico-Algebraic Approach to Quantum
Mechanics: Volume I: Historical Evolution, pp. 293--328. Springer Netherlands
(1975)

\bibitem{Kock1971:bilinearity}
Kock, A.: Bilinearity and cartesian closed monads.
\newblock Math. Scand. \textbf{29}, 161--174 (1971)

\bibitem{Lee1995:quasiprobinfinite}
Lee, H.W.: Theory and application of the quantum phase-space distribution
functions.
\newblock Phys. Rep. \textbf{259}(3), 147 -- 211 (1995)

\bibitem{MacLaneMoerdijk1992}
MacLane, S., Moerdijk, I.: Sheaves in Geometry and Logic: A First Introduction
to Topos Theory.
\newblock Springer, New York (1992)

\bibitem{Moggi1991}
Moggi, E.: Notions of computation and monads.
\newblock Inf. Comput. \textbf{93}(1), 55 -- 92 (1991)

\bibitem{Nuiten2011}
{Nuiten}, J.: {Bohrification of Local Nets of Observables}.
\newblock In: B.~{Jacobs}, P.~{Selinger}, B.~{Spitters} (eds.) Proceedings 8th
International Workshop on Quantum Physics and Logic, \emph{Electronic
	Proceedings in Theoretical Computer Science}, vol.~95, pp. 237--244. Open
Publishing Association (2012)

\bibitem{Hansen2013:extremalwitness}
{Ove Hansen}, L., {Hauge}, A., {Myrheim}, J., {{\O}yvind Sollid}, P.: Extremal
entanglement witnesses.
\newblock Int. J. Quantum Inf. \textbf{13}(08), 1550,060 (2015)

\bibitem{Petz1986}
Petz, D.: Sufficient subalgebras and the relative entropy of states of a von
neumann algebra.
\newblock Commun. Math. Phys. \textbf{105}(1), 123--131 (1986)

\bibitem{Petz1988}
Petz, D.: Sufficiency of channels over von neumann algerbas.
\newblock Q. J. Math. \textbf{39}(1), 97 (1988)

\bibitem{Picado2004:locales}
Picado, J., Pultr, A., Tozzi, A.: Locales.
\newblock In: M.~Pedicchio Cristina, W.~Tholen (eds.) Categorical Foundations -
Special Topics in Order, Topology, Algebra and Sheaf Theory,
\emph{Encyclopedia of Mathematics and its Applications}, vol.~97, pp.
49--101. Cambridge University Press, US (2004)

\bibitem{Raynaund2014}
Raynaud, G.: Fibred contextual quantum physics.
\newblock Ph.D. thesis, University of Birmingham (2014)

\bibitem{Rivas2014:markov}
Rivas, A., Huelga, S.F., Plenio, M.B.: Quantum non-markovianity:
characterization, quantification and detection.
\newblock Rep. Prog. Phys. \textbf{77}(9), 094,001 (2014)

\bibitem{RudolphWright2000}
Rudolph, O., Wright, J.D.M.: On unentangled {G}leason theorems for quantum
information theory.
\newblock Lett. Math. Phys. \textbf{52}(200), 239--245 (2000)

\bibitem{SpittersVickersWolters2014}
{Spitters}, B., {Vickers}, S., {Wolters}, S.: {Gelfand spectra in Grothendieck
	toposes using geometric mathematics}.
\newblock In: R.~{Duncan}, P.~{Panangaden} (eds.) {Proceedings 9th Workshop on
	Quantum Physics and Logic}, \emph{Electronic Proceedings in Theoretical
	Computer Science}, vol. 158, pp. 77--107. Open Publishing Association (2014)

\bibitem{TycVlach2015}
Tyc, T., Vlach, J.: Quantum marginal problems.
\newblock Eur. Phys. J. D \textbf{69}(9) (2015)

\bibitem{Vickers2007:localetoposasspaces}
Vickers, S.: Locales and toposes as spaces.
\newblock In: M.~Aiello, I.~Pratt-Hartmann, J.~Van~Benthem (eds.) Handbook of
Spatial Logics, pp. 429--496. Springer Netherlands (2007)

\bibitem{Vickers2011:monad}
Vickers, S.: A monad of valuation locales (2011).
\newblock Available at \url{http://www.cs.bham.ac.uk/~sjv/Riesz.pdf}

\bibitem{Wallach2002}
Wallach, N.R.: An unentangled {G}leason's theorem.
\newblock Contemp. Math. \textbf{305}, 291--298 (2002)

\bibitem{Westerbaan2017}
Westerbaan, A.: Quantum programs as kleisli maps.
\newblock In: R.~Duncan, C.~Heunen (eds.) {\rm Proceedings 13th International
	Conference on} Quantum Physics and Logic, \emph{Electronic Proceedings in
	Theoretical Computer Science}, vol. 236, pp. 215--228. Open Publishing
Association (2017)

\bibitem{Wolters2013}
Wolters, S.A.M.: A comparison of two topos-theoretic approaches to quantum
theory.
\newblock Commun. Math. Phys. \textbf{317}(1), 3--53 (2013)

\bibitem{WoltersHalvorson2013}
Wolters, S.A.M., Halvorson, H.: {Independence Conditions for Nets of Local
	Algebras as Sheaf Conditions}.
\newblock arXiv:1309.5639  (2013)

\end{thebibliography}
\end{document}